\documentclass[12pt]{article}
\usepackage{amsmath,amssymb,exscale}
\usepackage{mathrsfs}
\usepackage{graphicx}
\usepackage{hyperref}
\usepackage{blkarray}
\usepackage{cleveref}
\usepackage{enumerate}
\usepackage{cite}
\usepackage{tikz}
\usetikzlibrary{arrows,matrix,positioning}
\usepackage{enumitem}

\newcommand{\scp}[1]{\scriptstyle#1}
\DeclareMathOperator{\rank}{rank}
\input epsf
\allowdisplaybreaks
\newcommand\numberthis{\addtocounter{equation}{1}\tag{\theequation}}

\renewcommand\arraystretch{1.2}

\begin{document}
\topmargin -1.0cm
\oddsidemargin -0.8cm
\evensidemargin -0.8cm
\pagestyle{empty}
\vspace*{5mm}

\begin{center}

{\Large\bf The particle spectrum of  parity-violating Poincar\'e gravitational theory}

\vspace{1.0cm}

{\large  Georgios K. Karananas}\\

\vspace{.6cm}

 {\it {Institut de Th\'eorie des Ph\'enom\`enes Physiques, \'Ecole Polytechnique F\'ed\'erale de Lausanne,\\
CH-1015 Lausanne, Switzerland}}\\

\vspace{.6cm}
\texttt{georgios.karananas@epfl.ch}
\end{center}

\vspace{1cm}
\begin{abstract}
In this paper we investigate the physical spectrum of the gravitational theory based on the Poincar\'e group with terms which are at most quadratic in tetrad and spin connection, allowing for the presence of parity-even as well as parity-odd invariants. We determine restrictions on the parameters of the action so that all degrees of freedom propagate and are neither ghosts nor tachyons. We show that the addition of parity non-conserving invariants extends the healthy parameter space of the theory. To accomplish our goal, we apply the weak field approximation around flat spacetime and in order to facilitate the analysis, we separate the bilinear action for the excitations into completely independent spin sectors. For this purpose, we employ the spin-projection operator formalism and extend the original basis built previously, to be able to handle the parity-odd pieces.  
\end{abstract} 

\vfill
November 2014
\vfill

\eject
\pagestyle{empty}
\setcounter{page}{1}
\setcounter{footnote}{0}
\pagestyle{plain}

\section{Introduction}
\label{sec:Intro}

Gravitational interaction is usually treated in the context of Einstein's theory of General Relativity (GR), which has been very successful in the description of a plethora of macroscopic phenomena. However, gravitation is the least understood of interactions at microscopic scales, where quantum physics becomes important. In addition to that, the formulation of GR is rather different from the one of the Standard Model (SM) of particle physics that describes the electromagnetic, weak, and strong interactions. The latter is a theory built with the gauge principle of Weyl~\cite{Weyl:1929fm}, Yang and Mills~\cite{Yang:1954ek} as a guiding principle: the localisation of a symmetry group that initially acts rigidly is responsible for the emergence of an interaction. 

The pursuit of a gravitational theory with better microscopic behaviour that GR, as well as the fact that Yang-Mills theories enjoyed big success, initiated investigations~\cite{Utiyama:1956sy,Brodsky:1961} that eventually lead to the formulation of a gravitational theory that results from the gauging of the group of isometries (translations, Lorentz transformations) that acts rigidly on Minkowski spacetime, i.e. the Poincar\'e group~\cite{Sciama:1962,Kibble:1961ba}. Notice however that the resulting theory is not exactly GR, but an alteration that extends the Riemannian spacetime to a spacetime with curvature $R$ as well as torsion $T$.  

Once the standard gauging procedure is followed, the Poincar\'e group is made local by the introduction of two a priori independent gauge fields with appropriate transformation properties under group operations. The one associated to translations is the tetrad $e$ and the one associated to local Lorentz transformations is the spin connection $\omega$. In this setup, the gravitational interaction emerges from the gauging of the symmetry, in accordance with what happens in the SM.\footnote{Strickly speaking, this parallelism is not entirely true; the gauge symmetries of the SM are dictated by the groups $SU(3),SU(2) \ \text{and} \ U(1)$, which are all internal groups and hence do not affect spacetime. On the other hand, Poincar\'e group is an external group, i.e. it determines the spacetime symmetries. This however does not invalidate the gauging procedure.} 

As usual in the gauge theoretic approach, the covariant derivative is defined by using appropriately the connection $\omega$ and the field strengths of the group can be constructed in the standard way, i.e. by considering the commutator of two covariant derivatives. These are known in nomenclature as curvature $R$ and torsion $T$. With all ingredients at hand, the action describing the dynamics of the theory can be arranged in a systematic way by considering all possible invariants constructed from curvature and torsion at a given order in derivatives. We can schematically write
\begin{equation}
\label{introact}
\mathscr L = \mathscr L_0+\mathscr L_1(R,T)+\mathscr L_2(R,T)+\ldots \ ,
\end{equation}
where $\mathscr L_0, \mathscr L_1,\ldots,$ contain terms with zero derivatives (cosmological constant), one derivative (scalar curvature, Holst term) etc.. Notice that in this theory there are more degrees of freedom than in standard GR. At sufficiently low energies (below the masses of the particles), the heavy fields can be ``integrated out'', the equation of motion for the connection renders it non-dynamical and the Einstein-Hilbert action is recovered~\cite{Delacretaz:2014oxa}. If one requires the connection to be an independent propagating field, then terms beyond the leading order have to be taken into account.

The aim of this paper is to identify healthy subclasses of the Poincar\'e-invariant gravitational theory, with all possible parity-even as well as parity-odd terms that are at most quadratic in the field strengths $R$ and $T$.  
This clearly means that the action contains terms with \emph{two derivatives of the fields $e$ and $\omega$, at most}.  Let us explain why we restrict ourselves this way. From our point of view, the absence of terms with more than two derivatives is an essential requirement, since higher-derivative theories are usually plagued by ghosts. Since here tetrad and  connection are treated as independent fields, this theory should not be mistaken for a higher-derivative theory, but rather as ``gravity \`a la Yang-Mills''; this theory is dubbed Poincar\'e gauge theory of gravity (PGT) and it has been studied extensively in the literature~\cite{Neville:1978bk,Sezgin:1979zf,Hayashi:1980,Sezgin:1981xs,Neville:1981be,Nair:2008yh,Nikiforova:2009qr,Hernaski:2009wp}. An extensive review as well as historical details can be found for example in~\cite{Hehl:1976kj,Hehl:1994ue,Gronwald:1995em} and references therein. It is worth mentioning that PGT incorporates as simplest cases the Einstein-Cartan theory~\cite{Trautman:2006fp}, the teleparallel equivalent of GR~\cite{DeAndrade:2000sf}, as well as GR in the absence of fermionic matter. Given the fact that GR has been extremely successful in the description of Nature at large scales, the fact that PGT is capable of reducing to GR in certain limiting cases is encouraging. 

The most straightforward way to accomplish our goal is to determine the particle spectrum of the theory around the flat spacetime. In the present work, we do not discuss how the dynamics is modified when arbitrary curved spacetimes are considered as backgrounds, which constitutes a complicated problem that deserves to be adressed separately. It is well known that once a theory is studied around backgrounds different from the Minkowski one, especially if it contains massive spin-2 modes, pathologies might appear; this is what happens for example in the Fierz-Pauli theory (Boulware-Deser effect). Notice though that this is not the case for certain subclasses of the PGT we consider here, which remain free from ghosts and tachyons  when studied around maximally symmetric backgrounds~\cite{Nair:2008yh}.

Investigating the behaviour of the physical propagator, we find constraints on the parameters of the action so that the propagating degrees of freedom are neither ghosts, nor tachyons. We believe that the reason we choose to proceed this way is clear: the poles of the propagator correspond to the masses of the particles the theory contains, whereas the sign of the residues evaluated at the poles determine whether or not the theory is ghost-free~\cite{Neville:1978bk,Sezgin:1979zf,Schwinger:1970xc}. 

Let us give some more details on the methodology followed in this paper. First, we linearise the action around Minkowski spacetime and we retain only the bilinear in the fluctuations terms. We then employ the spin-projection operator formalism initially developed by Barnes~\cite{Barnes:1965} and Rivers~\cite{Rivers:1964}, see also Ref.~\cite{VanNieuwenhuizen:1973fi}. This framework is very powerful and ideally suited for such kind of problems, the main advantage being that the action for the excitations naturally breaks into independent spin sectors. Meanwhile, the coefficients of the expansion of the action in the projectors' basis can be conveniently arranged in matrices. This fact, together with the simple orthogonality relations the operators satisfy, makes the attainment of the propagator a straightforward exercise.

The inclusion of parity-odd terms, however, makes this exercise algebraically much more involved with respect to a number of interesting works that have appeared over a period of many years~\cite{Neville:1978bk,Sezgin:1979zf,Sezgin:1981xs,Neville:1981be,Hernaski:2009wp}. In these papers, the authors concentrated mainly on parity-even theories and studied in depth their particle dynamics. It is our purpose here to extend these works by including parity non-conserving invariants. We hope that by considering the effects of these terms in a systematic way could lead to new directions towards understanding questions that are of big significance in Cosmology, like the baryon asymmetry of the Universe~\cite{Sakharov:1967dj}.

There have been studies on PGT with parity-violating terms that are relevant to what we do in this paper. 
The first one is work that has been carried out by Kuhfuss and Nitch in the 80's~\cite{Kuhfuss:1986rb}. They studied the teleparallel equivalent of GR, a certain sub-category of PGT with vanishing curvature. They considered in addition to the three parity-even torsion terms, a parity-odd torsion term. Since tetrad is the only dynamical degree of freedom of this theory, they derived projectors associated to the tetrad perturbations only. The other one is a very interesting and relatively recent work by Hehl and collaborators~\cite{Baekler:2010fr}. In their paper, the authors allow for parity-odd pieces in a particular case of PGT that propagates only scalar degrees of freedom. This theory has interesting cosmological applications~\cite{cosmo-torsion} and it has been argued that it remains consistent in the non-linear level as well~\cite{Hamiltonian}. 
The authors determine necessary and sufficient conditions on the parameters of their theory so that it is physically acceptable. Notice that they did not resort to linearisation or the use of projection operators, but instead they partially diagonalised the initial Lagrangian for the case where spin-2 torsion vanishes. Finally, we would like to mention that there has been some renewed interest in three-dimensional PGT and especially on the effect of the gravitational Chern-Simons term, see for example~\cite{HelayelNeto:2010jn} and references therein.

The present paper is organised as follows. In Section~\ref{sec:Poincareg} we introduce the 14-parameter theory under investigation and we present the linearised quadratic action for the perturbations. In Section~\ref{sec:spinproj} we review the spin-projection formalism that is used to decompose the theory into independent spin sectors. Since we want to elucidate the role of parity-violating terms by treating them in the same footing as parity-preserving ones, we expand the original basis of projectors built in~\cite{Neville:1978bk,Sezgin:1979zf}, by introducing appropriate operators that allow us to work with terms that contain the totally antisymmetric tensor; most of them have never appeared before, as far as we know. In Section~\ref{sec:Particle Content} we find the constraints on the parameters of the action so that it propagates only healthy degrees of freedom. This we achieve by requiring positive masses and residues of the propagators when evaluated at the poles. In~Section~\ref{sec:conclus} we present the concluding remarks. The full set of projection operators, as well as the coefficient matrices and their inverses are given in Appendix A. Details on the derivation of the projectors can be found in Appendix B. An alternative method for the determination of healthy sub-classes of the theory under investigation is developed in Appendix C. 

In our notation, Greek letters $(\lambda,\mu,\nu,\ldots)$ are reserved for spacetime indices, whereas capital Latin letters $(A,B,C,\ldots)$ for flat indices. We use the Landau-Lifshitz signature for the Minkowski metric, $\eta_{AB}=\text{diag}(+,-,-,-)$. The convention for the totally antisymmetric tensor is $\epsilon^{0123}=-\epsilon_{0123}=1$. To keep the expressions as simple as possible we set $M_{\text{Planck}}=\hbar=c=1$.

\section{Poincar\'e gravity}
\label{sec:Poincareg}

The \emph{Poincar\'e group} is the semi-direct product of translations and Lorentz transformations. Its role in particle physics is fundamental and twofold. On one hand, it dictates the symmetries of the underlying Minkowski spacetime of Special Relativity. On the other hand, particle states in quantum field theories are classified according to the unitary irreducible representations of this particular group~\cite{Wigner:1939cj,Weinberg:1995mt}.

As we have already briefly mentioned in the Introduction, the idea behind PGT is the gauging of the Poincar\'e group. Despite the fact that by definition it affects spacetime symmetries, the gauge principle is applicable and simply amounts to promoting the 10 (constant) parameters of the group to depend arbitrarily on position. Invariance under local Poincar\'e transformations is achieved by the introduction of 40 independent fields: 16 of them are associated to translations and comprise the tetrad $e_\mu^{\ A}$, whereas the remaining 24 are associated to local Lorentz transformations and comprise the (spin) connection $\omega_\mu^{ \ AB}=-\omega_\mu^{\ BA}$. The gravitational interaction in this setup stems from both tetrad and connection and is a consequence of the gauging of the symmetry.\footnote{From the geometric point of view, the localisation of the Poincar\'e group deforms the underlying Minkowski structure of spacetime and as a result a new geometry emerges, the Riemann-Cartan geometry.}

The covariant derivative, as in any gauge theory, is defined as $D_\mu^{\ AB}=(\partial_\mu+\omega_\mu)^{AB},$ therefore the field strengths, torsion $T$ and curvature $R$, are readily obtained by considering the commutator of two covariant derivatives acting for example on a vector field. They are respectively given by 
\begin{align}
&T_{\mu\nu}^{\ \ A}=\partial_\nu e_\mu^{\ A}+\omega_\mu^{\ AB} e_{\nu B}-(\mu\leftrightarrow \nu) \ ,\\
&R_{\mu\nu}^{\ \ AB}=\partial_\nu \omega_\mu^{\ AB}+\omega_\mu^{\ AC}\omega_{\nu C}^{ \ \ \ B}-(\mu\leftrightarrow \nu)  \ .
\end{align}
For later convenience, we note that the above field strengths can be written in the tangent basis with the help of the tetrad\footnote{In the tangent basis the indices are manipulated with the Minkowski metric.}
\begin{align}
&T_{ABC}=e^\mu_{\ A}e^\nu_{\ B}\eta_{CD} T_{\mu\nu}^{\ \ D} \ , \ \ R_{ABCD}= e^\mu_{\ A}e^\nu_{\ B}\eta_{CE}\eta_{DF}R_{\mu\nu}^{\ \ EF} \ . 
\end{align}

The most general theory invariant under translations and local Lorentz transformations, with terms that are at most quadratic in $T$ and $R$, reads~\cite{Neville:1978bk,Sezgin:1979zf,Obukhov:1987tz,Diakonov:2011fs,Baekler:2011jt}
\begin{align}
\label{acti}
\mathscr L=&-\lambda R +\frac{1}{12}(4t_1+t_2+3\lambda)\,T_{ABC}\,T^{ABC}-\frac{1}{3}(t_1-2t_3+3\lambda)\,T_{AB}^{\ \ \ B}\,T^{AC}_{\ \ \ C} \nonumber\\
&-\frac{1}{6}(2t_1-t_2+3\lambda)\,T_{ABC}\,T^{BCA}-\frac{1}{12}(t_4+4t_5)\,\epsilon^{ABKL}\,T_{ABC}\,T_{KL}^{\ \ \ C}\nonumber\\
&-\frac{1}{3}(t_4-2t_5)\,\epsilon^{ABKL}\,T_{CAB}\,T^C_{\ \ KL}+\frac{1}{6}(2r_1+r_2)\,R_{ABCD}\,R^{ABCD}\nonumber\\
&+\frac{2}{3}(r_1-r_2)\,R_{ABCD}\,R^{ACBD}+\frac{1}{6}(2r_1+r_2-6r_3)\,R_{ABCD}\,R^{CDAB}\nonumber\\
&+(r_4+r_5)\,R_{AB}\, R^{AB} +(r_4-r_5)\,R_{AB}\,R^{BA}-\frac{1}{6}(r_6-r_8)\,\epsilon^{ABKL}R\,R_{ABKL}\nonumber\\
&-\frac{1}{8}(r_7+r_8)\,\epsilon^{ABKL}\,R_{ABCD}\,R_{KL}^{\ \ \ CD} +\frac{1}{4}(r_7-r_8)\,\epsilon^{ABKL}\,R_{ABCD}\,R^{CD}_{\ \ \ KL}  \ \ .
\end{align}
Here $\lambda,t_i,r_i$ are 14 arbitrary dimensionless constants and 
\begin{equation}
R_{AB}=e^\mu_{\ A}e^\nu_{\ C}\, \eta_{BC}\, R_{\mu\nu}^{\ \ CD} \ , \ \ R=e^\mu_{\ A}e^\nu_{\ B}R_{\mu\nu}^{\ \ AB} \ .
\end{equation}
We have allowed for parity-even ($\lambda,t_1,t_2,t_3,r_1,r_2,r_3,r_4,r_5$) as well as parity-odd ($t_4,t_5,r_6,r_7,r_8$) terms and we chose these peculiar combinations of coefficients because in this way the expressions that appear in the propagators simplify a lot. As it will turn out, these 5 new parity-violating parameters modify in a non-trivial way the conditions for the absence of ghost and tachyons. We will come back to this point in Section~\ref{sec:Particle Content}.

Some comments concerning our Lagrangian are in order at this point. First  of all, we have not written down a cosmological constant term; we want the field equations to admit Minkowski spacetime as solution. In addition to that, we have not included the following four terms 
\begin{equation}
\label{not-inc}
R(\omega(e))\ ,\ \ \epsilon^{ABCD}\, R_{ABCD} \  ,\ \ R^2 \ , \ \epsilon^{IJKL}\, R_{ABIJ}\, R^{AB}_{\ \ \ KL} \ ,
\end{equation}
with $\omega(e)$ the ``torsion-free'' connection, which in terms of the tetrad is given by 
\begin{equation}
\omega(e)\equiv-\omega_\mu^{\ AB}(e)=\frac{1}{2}\left[e^{\nu A}(\partial_\mu e_\nu^{\ B}-\partial_\nu e_\mu^{\ B})+e^{\nu A}e^{\lambda B}e_{\mu C}\partial_\lambda e_\nu^{\ C}-(A\leftrightarrow B)\right] \ .
\end{equation}
The first two of them can be related to $R$ and/or torsion squared terms by virtue of
\begin{equation}
\int d^4x\, e\,R(\omega(e))=\int d^4x\, \left[ e\, R-\frac{1}{4}\,T_{ABC}\,T^{ABC}+\frac{1}{2}\,T_{ABC}\,T^{BCA}+T_{AB}^{\ \ \ B}\,T^{AC}_{\ \ \ C}  \right]  \ ,
\end{equation}
and up to a total derivative
\begin{equation}
\int d^4 x\, e\, \epsilon^{ABKL}\,R_{ABKL}=-\frac{1}{2}\int d^4x \, e\, \epsilon^{ABKL}\,T_{ABC}T_{KL}^{\ \ \ C} \ .
\end{equation}
As for the $R^2$ term, it is related to $R_{AB}R^{BA}$ and $R_{ABCD}R^{CDAB}$, since for spaces topologically equivalent to flat, the Gauss-Bonnet theorem dictates
\begin{equation}
\int d^4x~e\Bigg[R^2-4\, R_{AB}\,R^{BA}+R_{ABCD}\,R^{CDAB}\Bigg]=0 \ .
\end{equation}
Finally, the term $\epsilon^{IJKL}\, R_{ABIJ}\, R^{AB}_{\ \ \ KL}$ does not need to be included since it is a total derivative.  

Before moving on, we would like to stress again that the PGT under consideration contains terms which are at most quadratic in the derivatives of the independent gauge fields $e_\mu^{\ A}$ and $\omega_\mu^{\ AB}$. Therefore, it should not be mistaken for a higher-derivative theory that usually suffer from unitarity issues. One notable exception is~\cite{Stelle:1977ry}
$$S=\int d^4x~e\Big[ R(\omega(e))+c R(\omega(e))^2 \Big]  \ , $$ 
with $c$ a positive constant. This theory in addition to the graviton, contains one healthy scalar degree of freedom and provides a viable inflationary model able to describe the Universe evolution in its primordial stages~\cite{Starobinsky:1980te}. 

We should also mention that in this paper we do not consider another very important class of modified gravity theories, namely the scalar-tensor theories, in which the gravitational degrees of freedom are contained into scalar field(s) and the metric. In a very interesting work carried out in the 70's by Horndeski~\cite{Horndeski:1974wa}, the most general  higher-derivative, but at the same time ghost-free, theory was constructed. The author showed that once appropriate couplings of a scalar field to the various curvature invariants (constructed from the metric only) are considered, even though the action of the theory contains more than two derivatives, the resulting equations of motion are at most second order. Therefore, the dynamics of the system described by this theory has in principle a well defined vacuum state (i.e. it is free from ghost instabilities). Recently, there has been renewed interest in Horndeski's theory, mainly in the context of its cosmological phenomenology, see for example~\cite{Horndeski-cosm} and references therein.

Let us now return to the theory under investigation and linearise the action~\eqref{acti} by considering the weak field approximation for the fields
\begin{equation}
\label{exp}
e_{\mu}^{\ A}\approx \eta_{\mu}^{\ A}+h_\mu^{\ A} \ ,\ h_{\mu}^{\ A}\ll 1 \ \ \ \text{and} \ \ \ \omega_{\mu}^{\ AB}\ll  1\ .
\end{equation}
In this limit there is no need to keep the distinction between spacetime and Lorentz indices, so in what follows we will use only capital Latin letters for tensorial quantities. It is also convenient to split the tetrad excitations into symmetric and antisymmetric parts, i.e. 
\begin{equation}
h_{AB}=s_{AB}+a_{AB}\ ,
\end{equation}
with 
\begin{equation}
s_{AB}=\frac{1}{2}(h_{AB}+h_{BA}) \ \ \ \text{and}\ \ \ a_{AB}=\frac{1}{2}(h_{AB}-h_{BA}) \ .
\end{equation}
Using the decomposition \eqref{exp} in the action, expanding in powers of $h_\mu^{\ A}$ and $\omega_{\mu}^{ \ AB}$ and retaining only the bilinear in perturbations parts, the linearised theory can be expressed as the sum of several terms that contain pure connection and tetrad excitations, as well as their mixings
\begin{equation}
\label{lin-terms}
S_2=S_2(\omega,\omega)+S_2(s,s)+S_2(a,a)+S_2(\omega,s)+S_2(\omega,a)+S_2(s,a) \ .
\end{equation}

A lengthy calculation reveals that each of the above terms reads
\begin{eqnarray}
S_2(\omega,\omega)&=&\displaystyle\frac{1}{12}\int d^4x~\Big\{4(2r_1-2r_2+3r_4+3r_5)\partial^{B}\omega_{CAB}\partial_D\omega^{CAD}+24(r_4+r_5)\partial^C\omega_B^{\ BA}\partial^D\omega_{CAD}\nonumber \\
& &\hspace{1.8cm}+3(r_7+r_8)\epsilon^{ABIJ}\partial_C\omega_{CAB}\partial^K\omega_{KIJ}-12(r_7-r_8)\epsilon^{ABIK}\partial_I\omega_{CAB}\partial_J\omega_K^{\ CJ} \nonumber  \\
& &\hspace{1.8cm}-3(r_7+r_8)\epsilon^{ABIJ}\partial_D\omega_{CAB}\partial^D\omega^{CIJ}-8(r_6-r_8)\epsilon^{ABCD}\partial_D\omega_{ABC}\partial_I\omega_K^{\ KI}  \nonumber \\
& &\hspace{1.8cm}+16(r_1-r_2)\partial^C\omega_{CAB}\partial_D\omega^{ABD}-4(2r_1+r_2)\partial^C\omega_{CAB}\partial_D\omega^{DAB} \nonumber \\
& &\hspace{1.8cm}+4(2r_1+r_2)\partial_D \omega_{CAB}\partial^D\omega^{CAB}+8(r_1-r_2)\partial_D \omega_{CAB}\partial^D\omega^{ACB}\nonumber \\
& &\hspace{1.8cm}+12(r_4+r_5)\partial_D\omega_B^{\ BA}\partial^D\omega^C_{\ CA}+12(r_4-r_5)\partial_A\omega_C^{\ CA}\partial^B\omega_D^{\ DB}
\nonumber \\
& &\hspace{1.8cm}+4(4r_1+2r_2-4r_3+3r_4-3r_5)\partial^{B}\omega_{CAB}\partial_D\omega^{ACD}\nonumber  \\
& &\hspace{1.8cm}-24t_5\epsilon^{ACIK}\omega_{CAB}\omega_{KI}^{\ \ B}-4(t_1-2t_3)\omega_B^{\ BA}\omega^C_{\ CA}\nonumber \\
& &\hspace{1.8cm}+4(t_1+t_2)\omega_{CAB}\omega^{CAB}-4(t_1-2t_2)\omega_{CAB}\omega^{ACB}\nonumber \\
& &\hspace{1.8cm}-8(t_4-2t_5)\epsilon^{ABIK}\omega_{CAB}(2\omega_{KI}^{\ \ B}+\omega^B_{\ KI})\Big\} \ , \\ 
\nonumber \\
S_2(s,s)&=&\displaystyle\frac{1}{3}\int d^4x~\Big\{3(t_1+\lambda)\partial_C s_{AB}\partial^C s^{AB} -(t_1-2t_3+3\lambda)\,(\partial_A s\partial^A s\nonumber\\
& &\hspace{1.8cm}-2\partial_A s\partial_B s^{AB} )-2(2t_1-t_3+3\lambda)\partial_B s_A^{\ B} \partial_C s^{AC}\Big\}
\ , \\
\nonumber \\
S_2(a,a)&=&\displaystyle\frac{1}{3}\int d^4x~\Big\{(t_1+t_2)\partial_C a_{AB}\partial^C a^{AB}-2(t_2-t_3)\partial_B a_A^{\ B}\partial_C a^{AC}\nonumber \\& &\hspace{0.6cm}+(t_4-2t_5)\epsilon^{ABKL}\left(\partial_C a_{AB}\partial^C a_{KL}-2\partial_C a_A^{\ C} \partial_L a_{BK}\right)\Big\}
\ , \\
\nonumber \\
S_2(\omega,s)&=&\displaystyle\frac{2}{3}\int d^4x \Big\{t_1\,\omega^{CAB}\partial_B s_{CA}+(t_1-2t_3)\omega_{C}^{\ CA}\left(\partial_B s_A^{\ B}-\partial_A s\right)\,\nonumber \\
& &\hspace{3.6cm}+2(t_4+t_5)\epsilon^{AKLM}\omega_{KLB}\partial_M s_A^{\ B} \nonumber \\
& &\hspace{3.9cm}+(t_4-2t_5)\epsilon^{AKLM}\omega_{BKL}\partial_M s_A^{\ B} \Big\} \ ,\\
\nonumber \\
S_2(\omega,a)&=&\displaystyle\frac{2}{3}\int d^4x\Big\{(t_1-2t_3)\omega_C^{\ CA}\partial_B a_A^{\ B}-(t_1-2t_2)\omega^{CAB}\partial_B a_{CA} \nonumber \\
& &\hspace{0.2cm}+(t_1+t_2)\omega^{ABC}\partial_A a_{BC}+6t_5\epsilon^{AKLM}\omega_{KLB}\partial_M a_A^{\ B} 
\nonumber \\
& &\hspace{2.2cm}+(t_4-2t_5)\epsilon^{ABKL}\left(\omega_{CKL}+\omega_{KLC}\right)\partial^C a_{AB}\nonumber \\
& &\hspace{3.5cm}-(t_4-2t_5)\epsilon^{AKLM}\omega_{BKL}\partial_M a_A^{\ B}\Big\} \ ,  \\
\nonumber \\
S_2(s,a)&=&\displaystyle\frac{2}{3}\int d^4x\Big\{2(t_1+t_3)\partial_B s_A^{\ B}\partial_C a^{AC}+(t_4-2t_5)\epsilon^{ABKL}\partial_C s_A^{\ C}\partial_B a_{KL}\Big\}\ .
\end{eqnarray}

After some straightforward algebraic manipulations that involve integration by parts, relabelling of indices etc., we recast the action into the following compact form\footnote{When convenient, we denote tensorial indices collectivelly by using Greek indices with acute accent $(\acute{\alpha},\acute{\beta},\ldots)$. This helps us to unclutter the notation and keep the expressions as short as possible.}
\begin{equation}
S_2=\frac{1}{2} \int d^4x \sum_{\acute{\alpha},\acute{\beta}} \phi_{\acute{\alpha}}~D_{\acute{\alpha}\acute{\beta}}~\phi_{\acute{\beta}} \ ,
\end{equation}
where the multiplet $\phi_{\acute{\alpha}}=(\omega_{CAB}, s_{AB}, a_{AB})$ contains the 40 components of the fields and the wave operator $D_{\acute{\alpha}\acute{\beta}}$ contains combinations of derivatives, the metric and the totally antisymmetric tensor. 

The quadratic action for the excitations~\eqref{lin-terms} has obviously inherited the linearised gauge symmetries of the original theory, i.e. it is invariant under 
\begin{equation}
\label{gaug-trans-1}
\delta h_{AB}=\partial_A \xi_B +\xi_{AB} \ ,\ \ \ \  \text{and}\ \ \ \  \delta \omega_{CAB}=-\partial_C \xi_{AB} \ ,
\end{equation}
where $\xi_A$ and $\xi_{AB}=-\xi_{BA}$ are the 10 gauge parameters of the Poincar\'e group. This fact has two important consequences. 

On one hand, since all fields appear with at most two derivatives in the action, it shows that 20 degrees of freedom are devoid of physical meaning and they can be set to zero by appropriately adjusting $\xi_A,\,\xi_{AB}$ and using the constraints. 
Therefore, out of the 40 independent fields we started with (16 in tetrad, 24 in connection), we are left with 20.\footnote{This is most easily seen in the canonical formalism, where the number of degrees of freedom is found by subtracting from the phase-space of the theory the number of constraints imposed by symmetries.} These are distributed among the different spin-sectors of the theory as follows: twelve are in the tensor part, which contains the massless graviton (two degrees of freedom) and two massive spin-2 fields (ten degrees of freedom). Six degrees of freedom are in the spin-1 part, which contains two massive vectors, whereas the remaining two comprise two massive scalar modes.

On the other hand, due to these symmetries, once we allow for the tetrad and connection to interact with appropriate external sources by introducing  
\begin{equation}
S_{sources}=\int d^4x~\Big[h^{AB}\,\tau_{AB}+\omega^{CAB}\,\sigma_{CAB} \Big] \ ,
\end{equation}
 we are immediately led to the following conservation laws 
\begin{equation}
\label{constr-1}
\partial^A \tau_{AB}= 0 \ , \ \ \ \ \text{and} \ \ \ \ \partial^C \sigma_{CAB}+\tau_{[AB]}= 0 \ .
\end{equation}
These 10 constraints on the sources will turn out to be very helpful in what follows.

\section{The spin-projection operator formalism}
\label{sec:spinproj}

In this section we lay the foundations in order to determine the spectrum of the theory in a systematic way. Our strategy is to study the behaviour of the (gauge-invariant)  \emph{saturated} propagator (i.e. the propagator sandwiched between conserved sources)
\begin{equation}
\label{prop1}
\Pi=-\sum_{\acute{\alpha},\acute{\beta}} j_{\acute{\alpha}}~D^{-1}_{\ \ \ \acute{\alpha}\acute{\beta}}~j_{\acute{\beta}} \ ,
\end{equation}
where the multiplet $j_{\acute{\alpha}}=(\sigma_{CAB}, \tau_{(AB)}, \tau_{[AB]})$ contains sources that couple only to the gauge-invariant components of the respective fields (physical sources).
We believe this is the most straightforward way to establish conditions on the parameters of the action, since the propagator contains all important information for the particle states predicted by the theory. First of all, the position of its poles correspond to the masses that have to be necessarily positive. Negative mass implies tachyonic behaviour. Also, the sign of the residues when evaluated at the poles determine whether or not the particles are ghosts. Negative residues correspond to negative contributions to the imaginary part of scattering amplitudes, which puts the unitarity of the theory under scrutiny.

In order to obtain the propagator, the wave operator has to be inverted and this is a rather non-trivial task. However, our goal is greatly facilitated when we take into account that tetrad and connection are reducible with respect to the three-dimensional rotations group. Therefore, they can be decomposed into subspaces of dimension $2J+1$ with definite spin $J$ and parity $P$.\footnote{Notice that this decomposition has nothing to do with the details of a theory. It simply follows from the construction of irreducible representations from tensorial quantities. Notice also that the classification of particle states according to their spin and parity has only meaning in the rest frame.} In the absence of parity-odd terms, the wave operator breaks into independent sectors that connect states with the same $J^P$ as follows:
\begin{center}
\begin{tabular}{|c|c|}
\hline
$J^P$ & sub-block dimension \\
\hline
$2^-$ & $1\times 1$ \\
\hline
$2^+$ & $2\times 2$ \\
\hline
$1^-$ & $4\times 4$ \\
\hline
$1^+$ & $3\times 3$ \\
\hline
$0^-$ & $1\times 1$ \\
\hline 
$0^+$ & $3\times 3$\\
\hline
\end{tabular}
\end{center}

To be able to proceed with this decomposition, it is very convenient to work in momentum space and employ the spin-projection operator formalism that was initially developed by Barnes~\cite{Barnes:1965} and Rivers~\cite{Rivers:1964}. The building blocks are the four-dimensional transverse and longitudinal projection operators; in momentum space these are respectively given by
\begin{equation}
\label{buildthetaom}
\Theta_{AB}=\eta_{AB}-\frac{k_A k_B}{k^2} \ \ \ \text{and} \ \ \ \Omega_{AB}=\frac{k_A k_B}{k^2} \ .
\end{equation}

In their seminal works, Neville~\cite{Neville:1978bk} and Sezgin-van Nieuwenhuizen~\cite{Sezgin:1979zf} studied the spectrum of the most general Poincar\'e-invariant theory with parity-even terms. To accomplish that, they used $\Theta$ and $\Omega$ to construct a covariant basis of projectors $P^{\phi\chi}_{ij}(J^P)_{\acute{\alpha}\acute{\beta}}$, which map between subspaces of fields $\phi,\chi$ with the same $J^P$. The lowercase Latin indices ($i,j,\dots$) denote the multiplicity of operators. This basis consists of 40 operators and is complete and orthogonal\footnote{Notice that the position of indices other than Lorentz ones is not important.}
\begin{eqnarray}
\label{comps}
&\displaystyle\sum_{\phi,i,J^P}P^{\phi\phi}_{ii}(J^P)_{\acute{\alpha}\acute{\beta}}=\mathbb I_{\acute{\alpha}\acute{\beta}} \ , &
\\
\vspace{.3cm}
\label{orths}
&P^{\phi \Sigma}_{i k}(I^P)_{\acute{\alpha}}^{\ \acute{\mu}}~P^{T \chi}_{l j}(J^Q)_{\acute{\nu}\acute{\beta}}=\delta_{\Sigma T}\delta_{IJ}\delta_{PQ}\delta_{kl}\delta^{\acute{\mu}}_{\acute{\nu}}P^{\phi\chi}_{ij}(J^P)_{\acute{\alpha}\acute{\beta}} \ .&
\end{eqnarray}

Let us move to the case of interest to us, i.e. the presence of parity-odd terms in the Lagrangian. The wave operator will now decompose into subspaces of same spin but not necessarily of same parity. A simple counting exercise reveals that the wave operator breaks into 3 independent spin sectors: one $3\times 3$ corresponding to spin-2 states, one $4\times 4$ corresponding to spin-0 states and a $7\times 7$ corresponding to spin-1 states.

It is obvious from the orthogonality conditions \eqref{orths} that the above-mentioned set of projectors is not able to handle the presence of terms that involve the totally antisymmetric tensor, since they cannot link states with same spin but different parity. It is therefore unavoidable to introduce new operators to take care of this; it turns out that in order to account for all possible mappings inside each spin sector, it is necessary to practically double in size the original basis built by Sezgin and van Nieuwenhuizen by adding 34 new operators. It is our understanding that this is the first time transition projectors that account for the parity-odd terms involving the connection is presented.\footnote{Kuhfuss and Nitsch~\cite{Kuhfuss:1986rb} introduced mixing projectors in order to study the interaction of states with different parity but only for the tertrad excitations.} 

 In our case, the completeness relation of eq. \eqref{comps} remains unchanged 
\begin{equation}
\label{compl}
\displaystyle\sum_{\phi,i,J}P^{\phi\phi}_{ii}(J)_{\acute{\alpha}\acute{\beta}}=\mathbb I_{\acute{\alpha}\acute{\beta}} \ , 
\end{equation}
whereas the orthogonality relation becomes 
\begin{equation}
\label{orth}
P^{\phi \Sigma}_{i k}(I)_{\acute{\alpha}}^{\ \acute{\mu}}~P^{T \chi}_{l j}(J)_{\acute{\nu}\acute{\beta}}=\delta_{\Sigma T}\delta_{IJ}\delta_{kl}\delta^{\acute{\mu}}_{\acute{\nu}}P^{\phi\chi}_{ij}(J)_{\acute{\alpha}\acute{\beta}} \ .
\end{equation}
Notice that we have suppressed the parity index. The full list of projectors as well as details on their derivation are given in the Appendix A and B respectively. 

In terms of the spin-projection operators, the action for the theory becomes 
\begin{equation}
S_2=\int d^4x \sum_{\phi,\chi,\acute{\alpha},\acute{\beta},i,j,J} c^{\phi\chi}_{ij}(J)~\phi_{\acute{\alpha}}~P^{\phi\chi}_{ij}(J)_{\acute{\alpha}\acute{\beta}}~\chi_{\acute{\beta}} \ ,
\end{equation}
where $c^{\phi\chi}_{ij}(J)$ are matrices that contain the coefficients of the expansion of the wave operator in the spin-projection operators basis. All ``physical information'' of the theory is contained in the $c^{\phi\chi}_{ij}(J)$ matrices: the zeros of their determinants correspond to the poles of the propagators, whereas their values at the poles correspond to the residues.

As we mentioned earlier, the action for the perturbations possesses certain gauge symmetries; namely it is invariant under the linearised form of general coordinate and local Lorentz transformations \eqref{gaug-trans-1}. These invariances manifest themselves in the spin-projectors language as well.  The way this happens is through degenerate coefficient matrices. Let us explain what this means. 

Assume that a matrix $M_{ij}(J)$ has dimension $(d\times d)$ and $\rank\left(M_{ij}(J)\right)=r$, so there exist $(d-r)$ right null eigenvectors $v_j^{R}(J)$ as well as $(d-r)$ left null eigenvectors $v_j^{L}(J)$. Consider the $n^{th}$ right null eigenvector $v_j^{(R,n)}(J)$ which satisfies
\begin{equation}
\sum_j M_{ij}(J)v_j^{(R,n)}(J)=0 \ .
\end{equation}
From the above we are led to the following gauge invariances
\begin{align}
\label{pro-gau-inv}
&\delta\phi_{\acute{\alpha}}=\sum_{J,i,\acute{\beta},n}v_i^{(R,n)}(J)P^{\phi\chi}_{ij}(J)_{\acute{\alpha}\acute{\beta}}f_{\acute{\beta}}(J) \ \ \ \ \text{for all $j$}  \ ,
\end{align}
with $f_{\acute{\alpha}}(J)$ an arbitrary element of the group.  On the other hand, for the $n^{th}$ left null eigenvector $v_j^{(L,n)}(J)$ we have
\begin{equation}
\sum_j v_j^{(L,n)}(J)M_{ji}(J)=0 \ ,
\end{equation}
and as result the sources are subject to the following constraints
\begin{align}
\label{pro-sour-con}
&\sum_{i,\acute{\beta}}v_i^{(L,n)}(J)P^{\phi\chi}_{ij}(J)_{\acute{\alpha}\acute{\beta}}S_{\acute{\beta}}=0 \ \ \ \ \text{for all $j$}  \ .
\end{align}

In the theory under consideration, the $7\times 7$ matrix that describes the sector associated to the vector perturbations of the theory is singular and of rank 4. In addition to that, the $4\times 4$ matrix for the spin-0 sector is also singular and of rank 3.
Using the explicit expressions for these matrices (given in Appendix A), a direct calculation reveals that eqs. \eqref{pro-gau-inv} and \eqref{pro-sour-con} respectively yield 
\begin{equation}
\delta h_{AB}=\partial_A \xi_B +\xi_{AB} \ , \ \ \delta \omega_{CAB}=-\partial_C \xi_{AB} \ , 
\end{equation}
and
\begin{equation}
\label{pro-sour-con-2}
\partial^A \tau_{AB}= 0 \ ,\ \ \partial^C \sigma_{CAB}+\tau_{[AB]}= 0 \ .
\end{equation}
The above result is expected and should not come as a surprise. 

At this point we can  proceed with the inversion of the coefficient matrices and calculate the propagator. In order to do so and since some of the $c^{\phi\chi}_{ij}(J)$ are singular, we simply have to invert the largest non-singular sub-matrix $b^{\phi\chi}_{ij}(J)$ extracted from them~\cite{Sezgin:1979zf,VanNieuwenhuizen:1973fi,Berends:1979rv}. Deleting $(d-r)$ rows and columns, practically amounts to imposing $(d-r)$ gauge conditions. Notice however that the gauge invariance of the propagator is guaranteed due to the $(d-r)$ source constraints given in~\eqref{pro-sour-con}. By virtue of the completeness and orthogonality relations~\eqref{compl} and~\eqref{orth} that $P^{\phi\chi}_{ij}(J)$ obey, the saturated propagator~\eqref{prop1} is given by
\begin{equation}
\label{propgafull}
\Pi=-\sum_{J,\phi,\chi,\acute{\alpha},\acute{\beta},i,j} \left(b^{ \phi\chi}_{ij}(J)\right)^{-1}j_{\acute{\alpha}}^*~P^{\phi\chi}_{ij}(J)_{\acute{\alpha}\acute{\beta}}~j_{\acute{\beta}} \ .
\end{equation}

\section{Particle Content}
\label{sec:Particle Content}

In this section we apply the formalism presented previously and we determine the restrictions on the parameters of the action~\eqref{acti}. 

\subsection{Massless sector}

Let us start in an unorthodox way by analysing first the massless sector of the theory. Since our result for the (massless) graviton must be proportional to the one that stems from Einstein's theory, this calculation provides a very useful check of our algebra. The projectors we use as a basis for expanding the wave operator are constructed with the use of $\Theta_{AB}$ and $\Omega_{AB}$ defined previously in~\eqref{buildthetaom}, as well as
\begin{equation}
\tilde k_A=\frac{k_A}{\sqrt{k^2}} \ .
\end{equation}
Subsequently, the limit $k^2=0$ has to be taken with some care. Apart from the genuine massless pole that corresponds to the graviton, we will also find $k^{-2n}~(n\ge 1)$ spurious singularities that originate from the operators and receive contributions from all spin sectors. Of course, the  propagator should be independent of the basis we use for the expansion. Therefore, all spurious singularities have to combine appropriately and cancel out in the final result, upon applying the source constraints. Since the expressions are rather involved and the calculations lengthy, we will omit them in what follows and we will only present the final results. The reader is referred to Appendix A for the explicit form of the coefficient matrices and the projection operators.

After a considerable amount of calculations involving all 74 projectors, we find that the cancellations between all spin sectors indeed take place in an elegant way and the residue of the propagator at the $k^2=0$ pole is 
\begin{equation} 
\text{Res}(\Pi; 0)=-\frac{1}{\lambda}\left(\partial_C\sigma^{ABC} \ \tau^{AB}  \right)\left(\begin{array}{ccc}
4&2\\
2&1
\end{array}\right)\left(\eta_{AI}\eta_{BJ}+\eta_{AJ}\eta_{BI}-\eta_{AB}\eta_{IJ}\right)\left(\begin{array}{c}
\partial_K\sigma^{IJK}\\
\tau^{IJ}
\end{array}\right) \ ,
\end{equation}
as it should. The requirement for absence of ghosts in the massless sector of the theory is therefore 
\begin{equation}
\lambda>0 \ .
\end{equation}

\subsection{Massive sector}

For massive states, the propagator for each spin sector can be written as
\begin{equation}
\label{propga}
\Pi(J)=-\frac{1}{(k^2-m_+(J)^2)(k^2-m_-(J)^2)}\sum_{\phi,\chi,\acute{\alpha},\acute{\beta},i,j} \left(b^{ \phi\chi}_{ij}(J)\right)^{-1}j_{\acute{\alpha}}~P^{\phi\chi}_{ij}(J)_{\acute{\alpha}\acute{\beta}}~j_{\acute{\beta}} \ ,
\end{equation}
by virtue of the completeness and orthogonality relations \eqref{compl} and \eqref{orth} that $P^{\phi\chi}_{ij}(J)$ obey. Here $b_{ij}^{\phi\chi}(J)$ is the residue matrix which is degenerate at the poles $k^2=m_\pm(J)^2$, with $m_\pm(J)$ the masses of the states. 
One might worry that the appearance of two poles in the propagator necessarily implies that one of the two states is ghost-like, since we can always write
\begin{equation}
\label{simp-fract}
\frac{1}{(k^2-m_+(J)^2)(k^2-m_-(J)^2)}=\frac{1}{m_+(J)^2-m_-(J)^2}\left(\frac{1}{k^2-m_+(J)^2}-\frac{1}{k^2-m_-(J)^2}\right) \ .
\end{equation}
However, this is not true here. The coefficient matrices contribute rather non-trivially to the residues and their values at one of the poles can differ significantly from their values at the other. 

The requirement for absence of tachyons and ghosts corresponds to real masses and positive-definite residues at the poles, i.e. 
\begin{align} 
\label{reqs}
&m_\pm(J)^2>0\ , \\
\label{reqs1}
&\displaystyle \sum_{i}\left[\left(b_{ii}^{\phi\chi}(J)\right)^{-1} P_{ii}^{\phi\chi}(J)\right]_{k^2=m_\pm(J)^2}>0 \ ,
\end{align}
where we suppressed tensorial indices in the diagonal projection operators. Since at the pole $P_{ii}^{\phi\chi}(J)$ contribute only a sign depending on the number of longitudinal operators $n_\Theta$ they contain, the condition~\eqref{reqs1} can be written equivalently as
\begin{equation}
\label{reqs2}
\sum_i (-1)^{n_\Theta}\left(b_{ii}^{\phi\chi}(J)\right)^{-1}_{k^2=m_\pm(J)^2}>0 \ .
\end{equation}

After a tedious calculation involving the coefficient matrices of the various spin sectors given in Appendix A, we apply~\eqref{reqs} and~\eqref{reqs2}, to find the following conditions on the parameters of the action for the absence of ghosts and tachyons in the massive sector of the theory
\begin{align*}
\label{conditions-full-s0}
\text{\hfill spin-0: \hfill}
&r_2<0\ ,\ \ \ 2r_2(r_1-r_3+2r_4)<-r_6^2\ ,\ \ \ r_1-r_3+2r_4>-\frac{r_6^2}{2r_2} \ ,\\
&\vphantom{\frac{r_6^2}{2r_2} }  t_2(t_3-\lambda)+t_4^2>0\ ,\ \ \ \left(t_2 t_3+t_4^2\right)\lambda(t_3-\lambda)>0 \numberthis
 \ ,\\
\\
\label{conditions-full-s1}
\text{\hfill spin-1: \hfill}
&(r_1+r_4+r_5)<0\ ,\ \ \  (r_1+r_4+r_5)(2r_3+r_5)<-r_7^2\ , \ \ \  2r_3+r_5>-\frac{r_7^2}{r_1+r_4+r_5}\ ,\\
&\vphantom{\frac{r_7^2}{r_1+r_4+r_5}}  (t_1+t_2)(t_1+t_3)+(t_4-2t_5)^2<0\ , \ \ \  t_2 t_3+t_4^2>0 \ , \\ &\vphantom{\frac{r_7^2}{r_1+r_4+r_5}}   t_1^2+4t_5^2>0\ ,\ \ \ t_3(t_1^2+4t_5^2)>-t_1(t_2t_3+t_4^2) \ , \numberthis \\ 
\\
\label{conditions-full-s2}
\text{\hfill spin-2: \hfill}
&r_1<0\ ,\ \ \ r_1(2r_1-2r_3+r_4)<-r_8^2\ , \ \ \ 2r_1-2r_3+r_4>-\frac{r_8^2}{r1} \ , \\
&\vphantom{\frac{r_8^2}{r1}}   t_1\lambda(t_1+\lambda)<0 \ , \ \ \ t_1(t_1+\lambda)+4t_5^2 >0 \ .\numberthis
\end{align*}

Let us now comment on our results. First of all, when parity-mixing terms are absent, the expressions above reduce exactly to the ones found by Sezgin-van Nieuwenhuizen~\cite{Sezgin:1979zf} and are presented below in~\eqref{conditions-sn-s0p}-\eqref{conditions-sn-s2m}. Meanwhile, it is apparent that the effect of the parameters corresponding to parity-odd invariants is indeed not-trivial: they are responsible for the fact that the inequallities we derived can be simultaneously satisfied. Take as an example the tensor part of the theory (eq.~\eqref{conditions-full-s2}). We see that if $t_5=0$, there is a contradiction, since the two constraints 
\begin{equation}
t_1\lambda(t_1+\lambda)<0 \ \ \ \text{and} \ \ \ t_1\lambda(t_1+\lambda)>0 \ ,
\end{equation}
cannot be simultaneously satisfied. Therefore, if we want healthy behaviour in the spin-2 sector of the PGT, we have two options: either we consider the most general case by imposing $t_5\neq 0$, or if we insist on restricting the parameter space by considering $t_5= 0$, we also have to set $t_1+\lambda=0$, or $r_1=0$, or $2r_1-2r_3+r_4=0$. This would correspond to getting rid of the massive  $2^-$ or $2^+$ field respectively, even though in the parity-violating theory we investigate, this distinction is not entirely accurate.\footnote{Strictly speaking, the massive states predicted by the theory are not parity eigenstates, due to the presence of parity-odd terms in the Lagrangian. However, we used the label $J^P$ for convenience.}

However, there is really no need to eliminate any of the massive poles in order to reconcile the inequalities~\eqref{conditions-full-s0}-\eqref{conditions-full-s2}, so we come to the rather unexpected conclusion that \emph{the most general quadratic in curvature and torsion gravitational theory based on the Poincar\'e group is ghost and tachyon free}. Notice that the designation ``most general'' corresponds to the PGT whose action contains all possible parity-conserving and parity-violating  invariants, which are at most quadratic in the derivatives of the gauge fields $e$ and $\omega$.

\subsection{Limiting cases}

Having determined the restrictions the parameters of the theory should obey, it is useful at this point to see what happens if we consider certain limiting cases in the PGT we study. Since this is the first time that an analysis on the full theory has been carried out, we believe that cross-checks on the results are crucial. First of all, once we consider parity-preserving invariants only, we recover the results of Sezgin-van Nieuwenhuizen~\cite{Sezgin:1979zf} that read
\begin{align*}
\label{conditions-sn-s0p}
&\text{\hfill spin-0$^+$: \hfill}
r_1-r_3+2r_4>0 \ ,\ \ \ t_3\lambda(t_3-\lambda)>0 \numberthis
 \ ,\\
\\
\label{conditions-sn-s0m}
&\text{\hfill spin-0$^-$: \hfill}
r_2<0 \ ,\ \ \ t_2>0 \numberthis
 \ ,\\
\\
&\text{\hfill spin-1$^+$: \hfill}
2r_3+r_5>0\ ,\ \ \ t_1 t_2(t_1+t_2)<0 \ ,\numberthis \\
\\
&\text{\hfill spin-1$^-$: \hfill}
r_1+r_4+r_5<0\ ,\ \ \ t_1 t_3(t_1+t_3)>0 \ ,\numberthis \\
\\
\label{conditions-sn-s2p}
&\text{\hfill spin-2$^+$: \hfill}
2r_1-2r_3+r_4>0\ , \ \ \ t_1\lambda(t_1+\lambda)<0 \ , \numberthis \\
\\
\label{conditions-sn-s2m}
&\text{\hfill spin-2$^-$: \hfill}
r_1<0\ , \ \ \ t_1>0 \ . \numberthis
\end{align*}
Of course, all 12 healthy subclasses of the above theory found in~\cite{Sezgin:1979zf} and~\cite{Sezgin:1981xs} are also limiting cases of the theory we consider here. To name a couple, if we keep only the term linear in curvature (this amount to setting in the above $t_i=0\ ,i=1,\ldots,5$ and $r_j=0\ ,j=1,\ldots,8$) we recover General Relativity. If we assume that there are no torsion terms present ($t_1=-t_2=-t_3=-\lambda$, $t_4=2t_5=0$), we find that the only acceptable theory is given by $r_2<0$ and $r_i=0\ , i=1,\ldots,8$. Notice that the coefficients of the parity-odd curvature terms have to be chosen equal to zero in order to avoid higher order poles in the propagators. Another interesting case is the teleparallel limit of the PGT given in~\eqref{acti},  studied in detail in~\cite{Kuhfuss:1986rb}. To consider this particular subclass, one has to impose vanishing curvature with an appropriate Lagrange multiplier. As a result, the only dynamical degrees of freedom are contained in the tetrad field. Since the coefficient matrices in this case are very simple, after a straightforward calculation one can reproduce the results of Kuhfuss and Nitsch.

So far we have been interested in the case of massive torsion exclusively. Since we have developed the necessary machinery, we can also consider the limit of massless torsion fields. This amounts to setting $t_i=0 \ , i=1,\ldots,5$ in the action~\eqref{acti}.\footnote{To be precise, this limit is not in accordance with the inequalities for the masses presented in~\eqref{conditions-full-s0}-\eqref{conditions-full-s2}, which we derived by assuming $t_i\neq 0,\ i=1,\ldots,5$. However, for the massless torsion case the only restrictions we find are the ones on the kinetic terms coefficients.}
On one hand, this case is somehow physically unattractive, since massless torsion leads to inconsistencies at the quantum level as presented extensively in~\cite{Shapiro:2001rz}. On the other hand, however, the propagators simplify significantly, since there is no (kinetic) mixing between tetrad and connection excitations. Consequently, the graviton propagator is sourced by tetrad only and has completely decoupled from the torsion propagator, that is sourced by the connection. 
This is apparent in the coefficient matrices presented in Appendix A, which they now assume block diagonal form. A short computation shows that the saturated propagator for this choice of parameters is naturally divided into two pieces as follows
\begin{equation}
\label{massless-torsion-limit}
\Pi=\Pi(h)+\Pi(\omega) \ . \\
\end{equation}
The first term $\Pi(h)$ corresponds to the graviton and is given by
\begin{equation}
\label{pih}
\Pi(h)=-\frac{1}{\lambda k^2}\left(\tau^*_{(AB)}\tau^{(AB)}-\frac{1}{2}\tau^*\tau\right) \ ,
\end{equation}
where we denoted $\tau_{(AB)}=\frac{1}{2}(\tau_{AB}+\tau_{BA})\ \text{and} \ \tau=\eta^{AB}\tau_{AB}$. 

The second term in eq.~\eqref{massless-torsion-limit} corresponds to torsion propagator and reads
\begin{align*}
\label{piom}
\Pi(\omega)=&-\frac{1}{k^2}\left\{\frac{2(r_1-r_3+2r_4)}{3(2r_2(r_1-r_3+2r_4)+r_6^2)}\left(\sigma^*_{CAB}\sigma^{CAB}-2\sigma^*_{CAB}\sigma^{CBA}\right)\right. -\frac{2r_1-2r_3+r_4}{r_1(2r_1-2r_3+r_4)+r_8^2}\times\\
&\hspace{-.4cm}\times\left.\left(\sigma^{*C}_{\ \ \ CB}\sigma_K^{\ \ KB}-\frac{2}{3}(\sigma^*_{CAB}\sigma^{CAB}-2\sigma^*_{CAB}\sigma^{CBA})\right)+\frac{2r_3+r_5}{(2r_3+r_5)(r_1+r_4+r_5)+r_7^2}\sigma^{*C}_{\ \ \ CB}\sigma_K^{\ \ KB} \vphantom{\frac{A}{B}}\right\}\\
&\hspace{-.4cm}+\frac{k^B k^J}{k^4}\left\{\frac{2}{3(2r_2(r_1-r_3+2r_4)+r_6^2)}\left(\vphantom{\frac{a}{b}}r_2\sigma^{*C}_{\ \ \ CB}\sigma^K_{\ \ KJ}-2(r_1-r_3+2r_4)\sigma^*_{CAB}(\sigma^{CA}_{\ \ \ J}-\sigma^{AC}_{\ \ \ J})\vphantom{\frac{a}{b}}\right)\right.\\
&\hspace{-.4cm}\left.+\frac{1}{3(r_1(2r_1-2r_3+r_4)+r_8^2)}\left(\vphantom{\frac{a}{b}}(4r_1-6r_3+3r_4)\sigma^{*C}_{\ \ \ CB}\sigma^K_{\ \ KJ}-(5r_1-8r_3+4r_4)\sigma^*_{CAB}\sigma^{CA}_{\ \ \ J}\right.\right.\\
&\hspace{-.4cm}\left.\left.-(r_1-4r_3+2r_4)\sigma^*_{CAB}\sigma^{AC}_{\ \ \ \ J}\vphantom{\frac{a}{b}}\right)-\frac{1}{(2r_3-r_5)(r_1+r_4+r_5)+r_7^2}\left(\vphantom{\frac{a}{b}}(2r_3-r_5)\sigma^{*C}_{\ \ \ CB}\sigma^K_{\ \ KJ}\right.\right.\\
&\hspace{-.4cm}\left.\left.-(r_1+r_4+r_5)\sigma^*_{CAB}\sigma^{CA}_{\ \ \ J}\vphantom{\frac{a}{b}}\right)\vphantom{\frac{A}{B}}\right\}-\frac{k_Ak^J}{k^4}\epsilon^{ABCI}\left\{\frac{r_6}{3(2r_2(r_1-r_3+2r_4)+r_6^2)}(\sigma^{*K}_{\ \ \ KJ}\sigma_{BCI}\right.\\
&\hspace{-.4cm}\left.+\sigma^*_{BCI}\sigma^{K}_{\ \ KJ})+\frac{r_7}{(2r_3+r_5)(r_1+r_4+r_5)+r_7^2}(\sigma^{*K}_{\ \ \ KB}\sigma_{CIJ}+\sigma^*_{CIJ}\sigma^{K}_{\ \ KB})\right. \\
&\hspace{-.4cm}\left. +\frac{r_8}{3(r_1(2r_1-2r_3+r_4)+r_8^2)}\left(\vphantom{\frac{a}{b}}(\sigma^*_{CIK}-\sigma^*_{KCI})(\sigma_{BJ}^{\ \ \ K}-\sigma^K_{\ \ BJ})\right.\right.\\
&\hspace{-.4cm}\left.\left.-(\sigma^{* \ \ K}_{BJ}-\sigma^{*K}_{\ \ \ BJ})(\sigma_{CIK}-\sigma_{KCI})\vphantom{\frac{a}{b}}\right)\vphantom{\frac{A}{B}}\right\} \numberthis \ .
\end{align*}

Several comments are in order concerning the above results for the propagators of the massless particles. As we already mentioned, because of the absence of kinetic mixing, the dynamics of the fields decouples completely. Therefore, the graviton exchange is due to the symmetric part of the tetrad source, as is apparent from eq.~\eqref{pih}. Notice that in the expression~\eqref{piom} apart from the standard well-behaved $k^{-2}$ poles, there are also $k^{-4}$ poles. The latter lead to states with energies not bounded from below, therefore they have to be eliminated. This can be easily achieved by tuning appropriately the coefficients appearing in front of these terms. However, unlike the massive case, the restrictions imposed this way are not enough to guarantee absence of ghosts~\cite{Sezgin:1981xs}. Let us illustrate why this is the case. Since we are now dealing with massless particles, we have to work in terms of helicity eigenstates by using the dyadic representation for the metric~\cite{Schwinger:1970xc}
\begin{equation}
\eta^{AB}=\frac{k^A\bar k^B+k^B\bar k^A}{k\bar k}+\sum_{\lambda=\pm 1}e^A_{k,\lambda}e^{*B}_{k,\lambda} \ ,
\end{equation}
where $\bar k^A$ is the time-reversed four-momentum and $e^A_{k,\lambda}$ satisfy
\begin{equation}
e^{*A}_{k,\lambda}=(-1)^\lambda e^A_{k,-\lambda}\ , \ \ \ e^{*A}_{k,\lambda}e_{A \ k,\lambda'}=\delta_{\lambda,\lambda'} \ ,\ \ \ k_A e^{A}_{k,\lambda}=\bar k_A e^{A}_{k,\lambda}=0 \ .
\end{equation} 
Upon plugging the above decomposition for the metric in the resulting propagator, we see that this choice cannot eliminate ghost terms proportional to $k\bar k$. For this to happen, we have to allow for appropriate gauge invariances in the theory, so that the source for the connection is conserved in the sense $k_B \sigma^{CAB}=0$. This can be achieved by inspecting the projectors and constructing degenerate coefficient matrices in a way similar to~\cite{Sezgin:1981xs}.

\section{Conclusions}
\label{sec:conclus}

In this paper we presented a systematic study of the spectrum of the most general gravitational theory that emerges from the gauging of the  Poincar\'e group. We considered terms that are at most quadratic in the field strengths and allowed for the presence of all possible parity-even as well as parity-odd invariants.  Our purpose was to fill a gap in previous analyses of Poincar\'e-invariant theories and demonstrate the influence of parity-violating terms in the dynamics of the particle states. 

We derived necessary and sufficient conditions on the 14 parameters of the action so that all spin sectors of the theory are free from ghosts and tachyons and propagate simultaneously. This was made possible by examining the behaviour of the (gauge-invariant) propagator when sandwiched between conserved sources for the tetrad and connection. After linearising the action around flat spacetime and moving to momentum space, we resorted to the spin-projection operator formalism that is used extensively for problems like the one addressed here. In order to account for terms that contain the totally antisymmetric tensor, we introduced in total 34 parity-violating projectors; most of them had never been constructed before. With the appropriate tools at hand we were able to decompose the action into 3 completely separate spin sectors and extract the corresponding coefficient matrices. Due to the presence of parity-odd terms, the computations concerning both massless and massive states was not as algebraically simple as in previous works. 

We considered first the massless sector of the theory that is a bit more involved in comparison to the massive one. Apart from the pole due to the graviton, the projection operators themselves introduce singularities at $k^2=0$. Since the choice of basis should not be of importance, we verified that these singularities are spurious and cancel in the final saturated propagator. We showed that the result for the graviton is identical to GR and at the same time we performed a non-trivial check of our algebra with this calculation.

We then turned our attention to the analysis of the massive degrees of freedom. Before inverting the coefficient matrices, we calculated the corresponding  determinants and specified what the physical masses of the particles are, i.e. where the poles of the propagators are located. Additionally, we found the residues of the propagators at the poles by inverting the coefficient matrices and evaluating them at the zeros of their determinants. 

Following that, we required: 
\vspace{-.3cm}
\begin{enumerate}[leftmargin=1.5cm]
\item Absence of negative masses, since they correspond to particles of tachyonic nature.

\item Positive-definite residues of the propagator at the poles; this guarantees that the particles' kinetic terms have the appropriate sign, therefore the theory is unitary.
\end{enumerate}

Imposing the above, we derived the constraints~\eqref{conditions-full-s0}-\eqref{conditions-full-s2} on the parameters of the theory, so that it contains only healthy states. As discussed in the main text, these inequalities can be satisfied simultaneously. Consequently, we can make the rather non-trivial statement that, apart from the massless graviton, the massive spin-2, spin-1 and spin-0 fields the theory contains, do not exhibit ghost-like or tachyonic behaviour. This is due to the inclusion of parity-violating invariants in the action.

We hope that with this paper we succeeded in filling the gap in the study of gravitational theories based on the Poincar\'e group. As we have shown, parity-odd terms are non-trivial modifications to the dynamics of the theory. Detailed analysis has to be made in order to see what the effects beyond the linear order are, or what happens when the theory is considered on backgrounds different from flat. 

The present work resulted from our attempt to built scale-invariant modifications of the PGT, i.e. a theory invariant under dilatations and thus, not containing any absolute energy scale. This we do by introducing a dilaton field in the action and coupling it appropriately to the various curvature and torsion invariants. We will use the formalism developed here and present our results in a future work~\cite{Karananas-ggravity2}.

\section*{Acknowledgements}
I am very grateful to Mikhail Shaposhnikov for suggesting the topic, numerous discussions and comments on the draft. I would also like to thank Friedrich Hehl, James Nester and Yuri Obukhov for correspondence. This work was supported by the Swiss National Science Foundation. 


\begin{appendix}

\renewcommand{\theequation}{A.\arabic{equation}}
\setcounter{equation}{0}

\section*{Appendix A}
\label{app:spinproj}

In this Appendix we first give the full set of spin-projection operators $P^{\phi\chi}_{ij}(J)_{\acute{\alpha}\acute{\beta}}$ that we used as a basis to break the theory into spin sub-blocks. We then present the coefficient matrices, as well as their inverses. We have arranged matters in such a way that the upper left sub-matrices always correspond to the negative parity states. When parity-violating terms are not present in the action, the matrices acquire block-diagonal form, so they can be inverted separately. This enables us to check our algebra easily by comparing with the results of Sezgin and van Nieuwenhuizen~\cite{Sezgin:1979zf}.
Finally, by looking at the zeros of the determinants, we write down the masses of the particles related to each spin sector.

In what follows, we denote with $\Theta_{AB}$ the transverse and with $\Omega_{AB}$ the longitudinal projection operators. In momentum space they are respectively given by
\begin{equation}
\Theta_{AB}=\eta_{AB}-\frac{k_A k_B}{k^2} \ , \ \ \ \text{and} \ \ \ \Omega_{AB}=\frac{k_A k_B}{k^2} \ .
\end{equation}
We also denote $\tilde k_A=k_A/\sqrt{k^2}$. It is understood that the projectors have to be symmetrised or antisymmetrised in their $(A,B)$ and $(I,J)$ indices, depending on the symmetries of the fields they act on. For example, $P^{\omega\omega}_{ij}(J)_{CABKIJ}$ have to be antisymmetrised in both $(A,B)$ and  $(I,J)$, whereas $P^{\omega s}_{ij}(J)_{CABIJ}$ have to be antisymmetrised in $(A,B)$ and symmetrised in $(I,J)$. 

The tensorial manipulations that are involved are quite tedious and prone to algebraic mistakes. For that reason, we have cross-checked extensively our calculations with \emph{Mathtensor}~\cite{Mathtensor:1994}.

\subsection*{A.1: Spin-0}

\renewcommand{\theequation}{A.1.\arabic{equation}}
\setcounter{equation}{0}

The 16 operators corresponding to the scalar part of the theory are
\begin{equation*}
\label{s0op}
\arraycolsep=.5cm\renewcommand\arraystretch{1.5}
\begin{array}{ll}
P^{\omega\omega}_{11}(0)_{CABKIJ}=\frac{1}{3} \Theta_{CK}\Theta_{AI}\Theta_{BJ}+\frac{2}{3} \Theta_{AK}\Theta_{BI}\Theta_{CJ} &P^{\omega s}_{23}(0)_{CABIJ}=\frac{\sqrt{2}}{3}\tilde k_B\Theta_{CA} \Theta_{IJ} \\
P^{\omega\omega}_{12}(0)_{CABKIJ}=\frac{1}{3}\epsilon_{ABCD}\Omega^D_I\Theta_{JK} &P^{s\omega }_{32}(0)_{ABKIJ}=\frac{\sqrt{2}}{3}\tilde k_J\Theta_{KI} \Theta_{AB} \\
P^{\omega\omega}_{21}(0)_{CABKIJ}=-\frac{1}{3}\epsilon_{IJKL}\Omega^L_A\Theta_{BC} &P^{\omega s}_{24}(0)_{CABIJ}=\sqrt{\frac{2}{3}}\tilde k_B \Theta_{CA} \Omega_{IJ} \\
P^{\omega\omega}_{22}(0)_{CABKIJ}=\frac{2}{3}\Theta_{BC}\Omega_{AI}\Theta_{JK} & P^{s\omega }_{42}(0)_{ABKIJ}=\sqrt{\frac{2}{3}}\tilde k_J\Theta_{K I} \Omega_{AB} \numberthis \\
P^{\omega s}_{13}(0)_{CABIJ}=\frac{1}{3\sqrt{2}}\epsilon_{ABCD}\tilde k^D \Theta_{IJ} &P^{ss}_{33}(0)_{ABIJ}=\frac{1}{3}\Theta_{AB}\Theta_{IJ}\\
P^{s\omega }_{31}(0)_{ABKIJ}=-\frac{1}{3\sqrt{2}}\epsilon_{IJKL}\tilde k^L \Theta_{AB}&P^{ss}_{34}(0)_{ABIJ}=\sqrt{\frac{1}{3}}\Theta_{AB}\Omega_{IJ} \\
P^{\omega s}_{14}(0)_{CABIJ}=\frac{1}{\sqrt{6}}\epsilon_{ABCD}\tilde k^D \Omega_{IJ}  &P^{ss}_{43}(0)_{ABIJ}=\sqrt{\frac{1}{3}}\Theta_{IJ}\Omega_{AB} \\
P^{s\omega }_{41}(0)_{ABKIJ}=-\frac{1}{\sqrt{6}}\epsilon_{IJKL}\tilde k^L \Omega_{AB} &P^{ss}_{44}(0)_{ABIJ}=\Omega_{AB}\Omega_{IJ} 
\end{array}
\end{equation*}

Using the above projectors we derived the $4\times 4$ coefficient matrix for the spin-0 sector that reads 
\begin{gather*}
\label{s0dm}
c_{ij}^{\phi\chi}(0)=\begin{blockarray}{ccccc}
\scp\omega^-&\scp\omega^+&\scp s^+&\scp s^+\\
\begin{block}{(cccc) c}
c_{11} & c_{12}  & c_{13} &c_{14} &\scp\omega^- \\
c_{21} & c_{22}  & c_{23} &c_{24} &\scp\omega^+ \\
c_{31} & c_{32}  & c_{33} &c_{34}&\scp s^+ \\
c_{41} & c_{42}  & c_{43} &c_{44}&\scp s^+ \\
\end{block} \numberthis
\end{blockarray}\ , \\
\begin{array}{llll}
c_{11}=k^2r_2+t_2 & c_{12}=k^2r_6-t_4 & c_{13}=-i\sqrt{2k^2}  t_4 & c_{14}=0 \\
c_{21}=-k^2r_6+t_4 & c_{22}=2k^2(r_1-r_3+2r_4)+t_3 & c_{23}=i\sqrt{2k^2}t_3 & c_{24}=0\\
c_{31}=-i\sqrt{2k^2} t_4 & c_{32}=-i\sqrt{2k^2}t_3 & c_{33}=2k^2(t_3-\lambda) & c_{34}=0\\
c_{41}=0 & c_{42}=0 & c_{43}=0 & c_{44}=0
\end{array}
\end{gather*}

Several comments concerning the above coefficient matrix are in order. First of all, the matrix is not Hermitian, something that can create confusion at first sight. This fact is simply a consequence of the normalisation of the corresponding parity-mixing projection operators. As discussed in detail in Appendix B, operators which connect the same states but contain the totally antisymmetric tensor, are required to have opposite signs. This is because we want them to obey the simple orthogonality relations given in eq.~\eqref{orth}, so that the inversion of the wave operator becomes straightforward. Obviously, the action is still Hermitian. 

In addition to that, the matrix~\eqref{s0dm} is clearly degenerate and of rank 3. This is expected due to the gauge invariances of the theory. To proceed with the attainment of the propagator we delete the last row and column of~\eqref{s0dm}. Denoting with $b_{ij}^{\phi\chi}(0)$ the resulting matrix, we perform the inversion to find
\begin{gather*}
\label{invs0dm}
\left(b_{ij}^{\phi\chi}(0)\right)^{-1}=\frac{k^2}{\det\left(b_{ij}^{\phi\chi}(0)\right)} \left(
\begin{array}{ccc}
B_{11}&B_{12}&B_{13}\\
B_{21}&B_{22}&B_{23}\\
B_{31}&B_{32}&B_{33}
\end{array} \right)\ , \numberthis \\ \\ 
\begin{aligned}
&\begin{array}{ll}
B_{11}=4k^2(r_1-r_3+2r_4)(t_3-\lambda)-2 t_3\lambda & B_{12}=-2k^2 r_6(t_3-\lambda)-2\lambda t_4 \\
B_{21}=2k^2 r_6(t_3-\lambda)+2\lambda t_4 & B_{22}=2k^2 r_2 (t_3-\lambda)+2\left(t_2(t_3-\lambda)+t_4^2\right)  \\
B_{31}=i\sqrt{2k^2}\left(r_6t_3+2(r_1-r_3-2r_4)t_4 \right) &B_{32}=i\sqrt{\frac{2}{k^2}}\left(k^2(r_2 t_3-r_6t_4)+t_2t_3+t_4^2\right)\\
&\\
\end{array} \\
&\begin{array}{l}
B_{13}=i\sqrt{2k^2}\left(r_6t_3+2(r_1-r_3-2r_4)t_4 \right)\\
B_{23}=-i\sqrt{\frac{2}{k^2}}\left(k^2(r_2 t_3-r_6t_4)+t_2t_3+t_4^2\right)\\
B_{33}=k^2\left(2r_2(r_1-r_3+2r_4)+r_6^2\right)+2(r_1-r_3+2r_4)t_2+r_2t_3-2r_6t_4+\frac{1}{k^2}\left(t_2t_3-t_4^2\right)\\
\end{array}
\end{aligned}
\end{gather*}

The determinant of the matrix can be written conveniently as
\begin{equation}
\det\left(b_{ij}^{\phi\chi}(0)\right)=2\left(2r_2(r_1-r_3+2r_4)+r_6^2\right)(t_3-\lambda)k^2(k^2-m_+(0)^2)(k^2-m_-(0)^2) \ ,
\end{equation}
where the masses of the spin-0 states $m_{\pm}(0)^2$, are given by
\begin{align*}
m_{\pm}(0)^2&=\frac{1}{2\left(2r_2(r_1-r_3+2r_4)+r_6^2\right)(t_3-\lambda)}\Bigg\{\left(2(r_1-r_3+2r_4)t_2+r_2 t_3-2r_6t_4\right)\lambda\\
&\hspace{-1cm}-2(r_1-r_3+2r_4)\left(t_2 t_3+t_4^2\right)
\pm \Bigg[4\left(2r_2(r_1-r_3+2r_4)+r_6^2\right)(t_2 t_3+t_4^2)(t_3-\lambda)\lambda \\
&\hspace{-1cm}+\left[\left(2(r_1-r_3+2r_4)t_2+r_2 t_3-2r_6t_4\right)\lambda-2(r_1-r_3+2r_4)\left(t_2 t_3+t_4^2\right)\right]^2 \Bigg]^{\frac{1}{2}} \Bigg\} \ .
\end{align*}
The notation we chose for the zeros of the determinant leaves no room for confusion; they correspond to the poles of the propagator, i.e. the physical masses of the spin-0 particle states of the theory. Therefore, they have to obey
\begin{equation}
m_+(0)^2 > 0 \ \ \ \text{and} \ \ \ m_-(0)^2 > 0 \ .
\end{equation}

In order to simplify as much as possible the calculations for the residue of the massless graviton, we found it helpful to isolate the $k^2=0$ pole in the spin-0 (and spin-2) sector of the theory. To do so, we rewrite the inverse of the coefficient matrix given above as
\begin{align}
\left(b_{ij}^{\phi\chi}(0)\right)^{-1}=&-\frac{1}{2\lambda k^2}\left(\begin{array}{ccc}
0&0&0\\
0&2k^2&-i \sqrt{2k^2}\\
0&i \sqrt{2k^2}&1
\end{array}\right)  \nonumber \\
\vphantom{\frac{a}{b}}\nonumber \\
&\hspace{-2cm}+\frac{1}{t_2\, t_3+t_4^2}\left(\begin{array}{ccc}
t_3&t_4&0\\
-t_4&t_2&0\\
0&0&-(2\lambda)^{-1}\Big(2(r_1-r_3+2r_4)t_2+r_2 t_3-2r_6t_4\Big)\end{array}\right)\nonumber \\
\vphantom{\frac{a}{b}}\nonumber \\
&\hspace{-2cm}+\frac{k^2}{2\big(2r_2(r_1-r_3+2r_4)+ r_6^2\big)(t_3-\lambda)\big(m_+(0)^2-m_-(0)^2\big)}\times \nonumber\\
&\times \left(\frac{1}{m_+(0)^2(k^2-m_+(0)^2)}-\frac{1}{m_-(0)^2(k^2-m_-(0)^2)}\right)
\left(\begin{array}{ccc}
B_{11}&B_{12}&B_{13}\\
B_{21}&B_{22}&B_{23}\\
B_{31}&B_{32}&B_{33}
\end{array} \right) \ ,
\end{align}
where the matrix elements $B_{ij}$ were given in~\eqref{invs0dm}.

\subsection*{A.2: Spin-1}

\renewcommand{\theequation}{A.2.\arabic{equation}}
\setcounter{equation}{0}

The 49 operators corresponding to the vector part of the theory are\\
\begin{equation*}
\label{s1op}
\arraycolsep=.6cm
\renewcommand\arraystretch{1.5}
\begin{array}{ll}
P^{\omega\omega}_{11}(1)_{CABKIJ}=\Theta_{CB}\Theta_{AI}\Theta_{JK} & P^{\omega a}_{24}(1)_{CABIJ}=2~\tilde k_B \Theta_{AI}\Omega_{CJ}  \\
P^{\omega\omega}_{12}(1)_{CABKIJ}=\sqrt{2}~\Theta_{CB}\Theta_{AI}\Omega_{JK} & P^{a\omega }_{42}(1)_{ABKIJ}=2~\tilde k_J \Theta_{AI}\Omega_{KB} \\
P^{\omega\omega}_{21}(1)_{CABKIJ}=\sqrt{2}~\Omega_{CB}\Theta_{AI}\Theta_{JK}&P^{\omega a}_{27}(1)_{CABIJ}=\epsilon_{AIJL}\tilde k^L\Omega_{BC}  \\
P^{\omega\omega}_{15}(1)_{CABKIJ}=\epsilon_{AJKL}\Omega_I^L\Theta_{BC}  & P^{a\omega }_{72}(1)_{ABKIJ}=-\epsilon_{IABD}\tilde k^D\Omega_{JK} \\
P^{\omega\omega}_{51}(1)_{CABKIJ}=-\epsilon_{IBCD}\Omega_A^D\Theta_{JK} &P^{\omega s}_{53}(1)_{CABIJ}=-\sqrt{2}\epsilon_{ACIL}\tilde k^L\Omega_{BJ} \\
P^{\omega\omega}_{16}(1)_{CABKIJ}=-\frac{1}{\sqrt{2}}~\epsilon_{AIJL}\Omega_K^L\Theta_{BC} &P^{s\omega }_{35}(1)_{ABKIJ}=\sqrt{2}\epsilon_{IKAD}\tilde k^D\Omega_{BJ} \\
P^{\omega\omega}_{61}(1)_{CABKIJ}=\frac{1}{\sqrt{2}}~\epsilon_{IABD}\Omega_C^D\Theta_{JK}
&P^{\omega a}_{54}(1)_{CABIJ}=-\sqrt{2}\epsilon_{ACIL}\tilde k^L\Omega_{BJ} \\
P^{\omega\omega}_{22}(1)_{CABKIJ}=2~\Omega_{CB}\Theta_{AI}\Omega_{JK}  & P^{a\omega }_{45}(1)_{ABKIJ}=\sqrt{2}\epsilon_{IKAD}\tilde k^D\Omega_{BJ}\\
P^{\omega\omega}_{25}(1)_{CABKIJ}=\sqrt{2}~\epsilon_{AJKL}\Omega_I^L\Omega_{BC} & P^{\omega a}_{57}(1)_{CABKIJ}=\sqrt{2}~\tilde k_B \Theta_{AI}\Theta_{CJ}  \\
P^{\omega\omega}_{52}(1)_{CABKIJ}=-\sqrt{2}~\epsilon_{IBCD}\Omega_A^D\Omega_{JK} & P^{a\omega}_{75}(1)_{CABKIJ}= \sqrt{2}~\tilde k_J \Theta_{AI}\Theta_{BK} \\
P^{\omega\omega}_{26}(1)_{CABKIJ}=-\epsilon_{AIJL}\Omega_K^L\Omega_{BC}
&P^{\omega s}_{63}(1)_{CABIJ}=-\epsilon_{ABIL}\tilde k^L\Omega_{CJ}  \\
P^{\omega\omega}_{62}(1)_{CABKIJ}=\epsilon_{IABD}\Omega_C^D\Theta_{JK}  & P^{s\omega }_{36}(1)_{ABKIJ}=\epsilon_{IJAD}\tilde k^D\Omega_{BK} \\
P^{\omega\omega}_{55}(1)_{CABKIJ}=\Theta_{CK}\Theta_{AI}\Omega_{BJ}+\Theta_{AK}\Omega_{BI}\Theta_{CJ}  & P^{\omega a}_{64}(1)_{CABIJ}=-\epsilon_{ABIL}\tilde k^L\Omega_{CJ} \numberthis \\
P^{\omega\omega}_{56}(1)_{CABKIJ}=-\sqrt{2}~\Omega_{AK}\Theta_{BI}\Theta_{CJ} &P^{a\omega }_{46}(1)_{ABKIJ}=\epsilon_{IJAD}\tilde k^D\Omega_{BK} \\
P^{\omega\omega}_{65}(1)_{CABKIJ}=-\sqrt{2}~\Omega_{CI}\Theta_{AJ}\Theta_{BK}
&P^{\omega a}_{67}(1)_{CABKIJ}= \tilde k_C \Theta_{AI}\Theta_{BJ} \\
P^{\omega\omega}_{66}(1)_{CABKIJ}= \Omega_{CK}\Theta_{AI}\Theta_{BJ}  & P^{a\omega}_{76}(1)_{CABKIJ}=\tilde k_K \Theta_{AI}\Theta_{BJ} \\
P^{\omega s}_{13}(1)_{CABIJ}=\sqrt{2}~\tilde k_J \Theta_{CB}\Theta_{AI} & P^{ss}_{33}(1)_{ABIJ}=2~\Theta_{AI}\Omega_{BJ}  \\
P^{s\omega }_{31}(1)_{ABKIJ}=\sqrt{2}~\tilde k_B \Theta_{KJ}\Theta_{AI}
& P^{sa}_{34}(1)_{ABIJ}=2~\Theta_{AI}\Omega_{BJ} \\
P^{\omega a}_{14}(1)_{CABIJ}=\sqrt{2}~\tilde k_J \Theta_{CB}\Theta_{AI} 
& P^{as}_{43}(1)_{ABIJ}=2~\Theta_{AI}\Omega_{BJ}  \\
P^{a\omega }_{41}(1)_{ABKIJ}=\sqrt{2}~\tilde k_B \Theta_{KJ}\Theta_{AI}
  & P^{s a}_{37}(1)_{ABIJ}=\epsilon_{AIJL}\Omega_{B}^L \\
P^{\omega a}_{17}(1)_{CABIJ}=\frac{1}{\sqrt{2}}\epsilon_{AIJL}\tilde k^L\Theta_{BC}
 & P^{a s }_{73}(1)_{ABIJ}=-\epsilon_{IABD}\Omega_{J}^D \\
P^{a\omega }_{71}(1)_{ABKIJ}=-\frac{1}{\sqrt{2}}\epsilon_{IABD}\tilde k^D\Theta_{JK}
 &P^{aa}_{44}(1)_{ABIJ}=2~\Theta_{AI}\Omega_{BJ}  \\
P^{\omega s}_{23}(1)_{CABIJ}= 2~\tilde k_B \Theta_{AI}\Omega_{CJ} 
&P^{a a}_{47}(1)_{ABIJ}=\epsilon_{AIJL}\Omega_{B}^L \\
P^{s\omega }_{32}(1)_{ABKIJ}=2~\tilde k_J \Theta_{AI}\Omega_{KB} 
 & P^{a a }_{74}(1)_{ABIJ}=-\epsilon_{IABD}\Omega_{J}^D\\
& P^{aa}_{77}(1)_{ABIJ}=\Theta_{AI}\Theta_{BJ} 
\end{array}
\end{equation*}

The $7\times 7$ coefficient matrix corresponding to spin-1 sector is found to be
\begin{gather*}
\label{s1dm}
c_{ij}^{\phi\chi}(1)=\begin{blockarray}{cccccccc}
\scp\omega^-&\scp\omega^-&\scp s^-&\scp a^-&\scp\omega^+&\scp\omega^+&\scp a^+ \\
\begin{block}{(ccccccc) c}
c_{11}&c_{12}&c_{13}&c_{14}&c_{15}&c_{16}&c_{17}&\scp\omega^- \\
c_{21}&c_{22}&c_{23}&c_{24}&c_{25}&c_{26}&c_{27}&\scp\omega^- \\
c_{31}&c_{32}&c_{33}&c_{34}&c_{35}&c_{36}&c_{37}&\scp s^- \\
c_{41}&c_{42}&c_{43}&c_{44}&c_{45}&c_{46}&c_{47}&\scp a^- \\
c_{51}&c_{52}&c_{53}&c_{54}&c_{55}&c_{56}&c_{57}&\scp \omega^+ \\
c_{61}&c_{62}&c_{63}&c_{64}&c_{65}&c_{66}&c_{67}&\scp \omega^+ \\
c_{71}&c_{72}&c_{73}&c_{74}&c_{75}&c_{76}&c_{77}&\scp a^+ \\
\end{block} 
\end{blockarray} \numberthis \\
\renewcommand\arraystretch{1.2}
\arraycolsep=.4cm
\begin{array}{lll}
c_{11}=k^2(r_1+r_4+r_5)+\frac{1}{6}(t_1+4t_3) &c_{12}=\frac{-1}{3\sqrt{2}}(t_1-2t_3)&c_{13}=\frac{i}{3}\sqrt{\frac{k^2}{2}}(t_1-2t_3) \\
c_{21}=\frac{-1}{3\sqrt{2}}(t_1-2t_3) &c_{22}=\frac{1}{3}(t_1+t_3)&c_{23}=-\frac{i}{3}\sqrt{k^2}(t_1+t_3) \\
c_{31}=-\frac{i}{3}\sqrt{\frac{k^2}{2}}(t_1-2t_3) &c_{32}=\frac{i}{3}\sqrt{k^2}(t_1+t_3)&c_{33}=\frac{1}{3}k^2(t_1+t_3) \\
c_{41}=-\frac{i}{3}\sqrt{\frac{k^2}{2}}(t_1-2t_3)&c_{42}=\frac{i}{3}\sqrt{k^2}(t_1+t_3)&c_{43}= \frac{1}{3}k^2(t_1+t_3)\\
c_{51}=k^2r_7-\frac{1}{3}(2t_4-t_5)&c_{52}=-\frac{\sqrt{2}}{3}(t_4-2t_5)&c_{53}=\frac{i}{3}\sqrt{2k^2}(t_4+t_5) \\
c_{61}=\frac{\sqrt{2}}{3}(t_4+t_5)&c_{62}=\frac{1}{3}(t_4-2t_5)&c_{63}=-\frac{i}{3}\sqrt{k^2}(t_4-2t_5) \\
c_{71}=\frac{i}{3}\sqrt{2k^2}(t_4+t_5)&c_{72}=\frac{i}{3}\sqrt{k^2}(t_4-2t_5)&c_{73}=\frac{1}{3}k^2(t_4-2t_5)  \\[.6cm]
c_{14}=\frac{i}{3}\sqrt{\frac{k^2}{2}}(t_1-2t_3) &c_{15}=-k^2r_7+\frac{1}{3}(2t_4-t_5)&c_{16}=-\frac{\sqrt{2}}{3}(t_4+t_5) \\
c_{24}=-\frac{i}{3}\sqrt{k^2}(t_1+t_3)&c_{25}=\frac{\sqrt{2}}{3}(t_4-2t_5)&c_{26}=-\frac{1}{3}(t_4-2t_5) \\
c_{34}=\frac{1}{3}k^2(t_1+t_3)  &c_{35}=\frac{i}{3}\sqrt{2k^2}(t_4+t_5)&c_{36}=-\frac{i}{3}\sqrt{k^2}(t_4-2t_5)  \\
c_{44}=\frac{1}{3}k^2(t_1+t_3) &c_{45}=\frac{i}{3}\sqrt{2k^2}(t_4+t_5)&c_{46}= \frac{i}{3}\sqrt{k^2}(t_4-2t_5) \\
c_{54}=\frac{i}{3}\sqrt{2k^2}(t_4+t_5)&c_{55}=k^2(2r_3+r_5)+\frac{1}{6}(t_1+4t_2)&c_{56}=\frac{1}{3\sqrt{2}}(t_1-2t_2) \\
c_{64}=-\frac{i}{3}\sqrt{k^2}(t_4-2t_5)&c_{65}=\frac{1}{3\sqrt{2}}(t_1-2t_2)&c_{66}=\frac{1}{3}(t_1+t_2) \\
c_{74}=\frac{1}{3}k^2(t_4-2t_5)&c_{75}=\frac{i}{3}\sqrt{\frac{k^2}{2}}(t_1-2t_2)&c_{76}=\frac{i}{3}\sqrt{k^2}(t_1+t_2)  \\[.6cm]
c_{17}=\frac{i}{3}\sqrt{2k^2}(t_4+t_5) &&\\
c_{27}=\frac{i}{3}\sqrt{k^2}(t_4-2t_5)&&\\
c_{37}=-\frac{1}{3}k^2(t_4-2t_5) &&\\
c_{47}=-\frac{1}{3}k^2(t_4-2t_5) &&\\
c_{57}=-\frac{i}{3}\sqrt{\frac{k^2}{2}}(t_1-2t_2) &&\\
c_{67}=-\frac{i}{3}\sqrt{k^2}(t_1+t_2) &&\\
c_{77}=\frac{1}{3}k^2(t_1+t_2) &&\\
\end{array}
\end{gather*}

As was the case in the spin-0 sector, the above matrix is not Hermitian because of the normalisation of the projectors that connect states with different parity. Also, due to the gauge invariances of the theory we expect this matrix to be singular. It turns out that the rank of the largest non-degenerate sub-matrix extracted from~\eqref{s1dm} is actually 4. We consider only the coefficients associated to connection excitations by dropping rows (and columns) 3, 6 and 7. We work with this particular sub-matrix purely for convenience. Clearly, this is not a unique choice. However, the propagator does not depend on what (regular) sub-matrix of rank 4 we study; its gauge invariance is guaranteed from the source constraints that we obtain.

To avoid confusion with the spin-0 sector, we denote the resulting matrix with $\widetilde b_{ij}^{\phi\chi}(1)$. It reads
\begin{gather*}
\label{s1dm}
\widetilde b_{ij}^{\phi\chi}(1)= \left(
\begin{array}{cccc}
\widetilde b_{11}&\widetilde b_{12}&\widetilde b_{13}&\widetilde b_{14}\\
\widetilde b_{21}&\widetilde b_{22}&\widetilde b_{23}&\widetilde b_{24}\\
\widetilde b_{31}&\widetilde b_{32}&\widetilde b_{33}&\widetilde b_{34}\\
\widetilde b_{41}&\widetilde b_{42}&\widetilde b_{43}&\widetilde b_{44}\\
\end{array} \right)\ , \numberthis \\ \\ 
\renewcommand\arraystretch{1.2}
\arraycolsep=.4cm
\begin{array}{lll}
\widetilde b_{11}=k^2(r_1+r_4+r_5)+\frac{1}{6}(t_1+4t_3) &\widetilde b_{12}=\frac{-1}{3\sqrt{2}}(t_1-2t_3)&\widetilde b_{13}=-k^2r_7+\frac{1}{3}(2t_4-t_5) \\
\widetilde b_{21}=\frac{-1}{3\sqrt{2}}(t_1-2t_3)&\widetilde b_{22}=\frac{1}{3}(t_1+t_3)& \widetilde b_{23}=\frac{\sqrt{2}}{3}(t_4-2t_5) \\
\widetilde b_{31}=k^2r_7-\frac{1}{3}(2t_4-t_5) &\widetilde b_{32}=-\frac{\sqrt{2}}{3}(t_4-2t_5) &\widetilde b_{33}=k^2(2r_3+r_5)+\frac{1}{6}(t_1+4t_2)\\
\widetilde b_{41}=\frac{\sqrt{2}}{3}(t_4+t_5) &\widetilde b_{42}=\frac{1}{3}(t_4-2t_5) &\widetilde b_{43}=\frac{1}{3\sqrt{2}}(t_1-2t_2)\\[.6cm]
\widetilde b_{14}=-\frac{\sqrt{2}}{3}(t_4+t_5)& &\\
\widetilde b_{24}=\frac{1}{3}(t_4-2t_5)& &\\
\widetilde b_{34}=\frac{1}{3\sqrt{2}}(t_1-2t_2)& &\\
\widetilde b_{44}=\frac{1}{3}(t_1+t_2)& &\\
\end{array} \\
\end{gather*}
The inverse of the above matrix can be written as
\begin{equation}
\label{invs1dm}
\left(\widetilde b_{ij}^{\phi\chi}(1)\right)^{-1}=\frac{1}{\det\left(\widetilde b_{ij}^{\phi\chi}(1)\right)} \text{adj}\left(\widetilde b_{ij}^{\phi\chi}(1)\right) \ ,
\end{equation}
where $ \text{adj}\left(\widetilde b_{ij}^{\phi\chi}(1)\right)$ is the adjoint of matrix~\eqref{s1dm}, whose elements are found to be
\begin{equation*}
\text{adj}\left(\widetilde b_{ij}^{\phi\chi}(1)\right)=\left(\begin{array}{cccc}
\widetilde B_{11}&\widetilde B_{12}&\widetilde B_{13}&\widetilde B_{14}\\
\widetilde B_{21}&\widetilde B_{22}&\widetilde B_{23}&\widetilde B_{24}\\
\widetilde B_{31}&\widetilde B_{32}&\widetilde B_{33}&\widetilde B_{34}\\
\widetilde B_{41}&\widetilde B_{42}&\widetilde B_{43}&\widetilde B_{44}\\
\end{array} \right)\ , \numberthis \\
\end{equation*}
\begin{equation*} 
\renewcommand\arraystretch{1.3}
\arraycolsep=.3cm
\begin{array}{l}
\displaystyle
\widetilde B_{11}=\frac{1}{18}\left\{\vphantom{\frac{a}{b}}2k^2(2r_3+r_5)\left((t_1+t_2)(t_1+t_3)+(t_4-2t_5)^2\right)+3\left(t_1^2 t_2+t_1(t_2 t_3+t_4^2)+4t_2t_5^2\right) \right\} \\
\begin{aligned}
\widetilde B_{21}=
\frac{1}{18\sqrt{2}}&\left\{\vphantom{\frac{a}{b}}2k^2\left[(2r_3+r_5)\left((t_1+t_2)(t_1-2t_3)-2(t_4-2t_5)(t_4+t_5)\right)\right.\right.\\
&\left.\left.+3r_7(t_1t_4+2t_2t_5)\right]+3\left(t_1^2t_2-2t_1(t_2t_3+t_4^2)+4t_2t_5^2\right)\vphantom{\frac{a}{b}} \right\}
\end{aligned}\\
\displaystyle
\widetilde B_{31}=-\frac{1}{18}\left\{\vphantom{\frac{a}{b}}2k^2 r_7\left((t_1+t_2)(t_1+t_3)+(t_4-2t_5)^2\right)-3\left[t_1^2t_4-2t_2\right] \right\} \\
\begin{aligned}
\widetilde B_{41}=-\frac{1}{18\sqrt{2}}&\left\{ \vphantom{\frac{a}{b}}
2k^2\left[ 
3(2r_3+r_5) \left(t_1t_4+2t_3t_5 \right)-r_7\left((t_1-2t_2)(t_1+t_3)\right.\right.\right.\\
&\left.\left.\left.-2(t_4-2t_5)(t_4+t_5)\right)\right]
+3t_1^2t_4+12\left(t_2t_3+(t_4+t_5)t_4\right)t_5
\vphantom{\frac{a}{b}}\right\}
\end{aligned}\\
\begin{aligned}
\widetilde B_{12}=\frac{1}{18\sqrt{2}}&\left\{\vphantom{\frac{a}{b}}2k^2\left[(2r_3+r_5)\left((t_1+t_2)(t_1-2t_3)-2(t_4-2t_5)(t_4+t_5)\right)\right.\right.\\
&\left.\left.+3r_7(t_1t_4+2t_2t_5)\right]+3\left(t_1^2t_2-2t_1(t_2t_3+t_4^2)+4t_2t_5^2\right) \vphantom{\frac{a}{b}}\right\}
\end{aligned}\\
\begin{aligned}
\widetilde B_{22}=\frac{1}{36}&\left\{\vphantom{\frac{a}{b}}12k^4\left[(2r_3+r_5)(r_1+r_4+r_5)+r_7^2)(t_1+t_2)\right]+2k^2\left[r_5\left((t_1+10t_2)t_1\right.\right.\right.\vphantom{\frac{a}{b}}\\
&\left.\left.\left.+4(t_1+t_2)t_3+4(t_4+t_5)^2\right)+2r_3\left((t_1+t_2)(t_1+4t_3)+4(t_4+t_5)^2\right)\right.\right.\vphantom{\frac{a}{b}}\\
&\left.\left.+3\left(3(r_1+r_4)t_1t_2-4r_7(t_4t_1-t_5t_2)\right)\right]\right.\vphantom{\frac{a}{b}}\\ 
&\left.+3\left[ t_1^2t_2+4(t_2t_3+t_4^2)+4t_2t_5\right]\vphantom{\frac{a}{b}}\right\}
\end{aligned}\\
\begin{aligned}
\widetilde B_{32}=\frac{1}{18\sqrt{2}}&\left\{\vphantom{\frac{a}{b}}2k^2\left[3(r_1+r_4+r_5)(t_1t_4+2t_2t_5)-r_7\left((t_1+t_2)(t_1-2t_3)\right.\right. \right.\\
&\left.\left.\left.-2(t_4-2t_5)(t_4+t_5)\right)\right]+3\left(t_1^2t_4+4(t_2t_3+(t_4+t_5)t_4)\right)t_5\vphantom{\frac{a}{b}}\right\}
\end{aligned}\\
\begin{aligned}
\widetilde B_{42}=\frac{1}{36}&\left\{-\vphantom{\frac{a}{b}}12k^4\left[\left((2r_3+r_5)(r_1+r_4+r_5)+r_7^2\right)(t_4-2t_5)\right]+2k^2\left[3\left[-(r_1+2r_3+r_4+2r_5)t_1t_4\right.\right.\right.\\
&\left.\left.\left.+4(r_1+r_3+r_5)t_2t_5+4(2r_3+r_5)t_3t_5\right]+r_7\left(t_1^2-2t_1(t_2+t_3)+4(t_2t_3+t_4^2\right.\right.\right.\\
&\left.\left.-7t_4t_5+t_5^2+3(t_4+4t_5))\right]\right\}
\end{aligned}\\ 
\displaystyle
\widetilde B_{13}=\frac{1}{18}\left\{\vphantom{\frac{a}{b}}2k^2 r_7\left((t_1+t_2)(t_1+t_3)+(t_4-2t_5)^2\right)-3\left[t_1^2t_4-2\left(t_2\right)\right] \right\} \\
\begin{aligned}
\widetilde B_{23}=-\frac{1}{18\sqrt{2}}&\left\{\vphantom{\frac{a}{b}}2k^2\left[3(r_1+r_4+r_5)(t_1t_4+2t_2t_5)-r_7\left((t_1+t_2)(t_1-2t_3)\right.\right. \right.\\
&\left.\left.\left.-2(t_4-2t_5)(t_4+t_5)\right)\right]+3\left(t_1^2t_4+4(t_2t_3+(t_4+t_5)t_4)\right)t_5\vphantom{\frac{a}{b}}\right\}
\end{aligned}\\
\begin{aligned}
\widetilde B_{33}=\frac{1}{18}&\left\{\vphantom{\frac{a}{b}}2k^2\left[(r_1+r_4+r_5)\left((t_1+t_2)(t_1+t_3)+(t_4-2t_5)^2\right)\right]\right.\\
&\left.+3\left(t_1^2t_3+t_1\left(t_2t_3+t_4^2\right)+4t_3t_5^2\right)\vphantom{\frac{a}{b}}\right\}
\end{aligned}\\
\begin{aligned}
\widetilde B_{43}=\frac{1}{18\sqrt{2}}&\left\{\vphantom{\frac{a}{b}}2k^2\left[(2r_1+2r_4-r_5)\left((t_1-2t_2)(t_1+t_3)-2(t_4-2t_5)(t_4+t_5)\right)\right.\right.\\
&\left.\left. -6r_7\left(t_1t_4+2t_3t_5\right)\right]+3\left(t_1^2t_3-2t_1(t_2t_3+t_4^2)+4t_3t_5^2\right)\vphantom{\frac{a}{b}}\right\}
\end{aligned}
\end{array}
\end{equation*}
\newpage
\begin{equation*}
\renewcommand\arraystretch{1.5}
\arraycolsep=.3cm
\begin{array}{l}
\begin{aligned}
\widetilde B_{14}=\frac{1}{18\sqrt{2}}&\left\{ \vphantom{\frac{a}{b}}
2k^2\left[ 
3(2r_3+r_5) \left(t_1t_4+2t_3t_5 \right)-r_7\left((t_1-2t_2)(t_1+t_3)\right.\right.\right.\\
&\left.\left.\left.-2(t_4-2t_5)(t_4+t_5)\right)\right]
+3t_1^2t_4+12\left(t_2t_3+(t_4+t_5)t_4\right)t_5
\vphantom{\frac{a}{b}}\right\}
\end{aligned}\\
\begin{aligned}
\widetilde B_{24}=-\frac{1}{36}&\left\{-\vphantom{\frac{a}{b}}12k^4\left[\left((2r_3+r_5)(r_1+r_4+r_5)+r_7^2\right)(t_4-2t_5)\right]+2k^2\left[3\left[-(r_1+2r_3+r_4+2r_5)t_1t_4\right.\right.\right.\\
&\left.\left.\left.+4(r_1+r_3+r_5)t_2t_5+4(2r_3+r_5)t_3t_5\right]+r_7\left(t_1^2-2t_1(t_2+t_3)+4(t_2t_3+t_4^2\right.\right.\right.\\
&\left.\left.-7t_4t_5+t_5^2+3(t_4+4t_5))\right]\vphantom{\frac{a}{b}}\right\}
\end{aligned}\\
\begin{aligned}
\widetilde B_{34}=\frac{1}{18\sqrt{2}}&\left\{\vphantom{\frac{a}{b}}2k^2\left[(2r_1+2r_4-r_5)\left((t_1-2t_2)(t_1+t_3)-2(t_4-2t_5)(t_4+t_5)\right)\right.\right.\\
&\left.\left. -6r_7\left(t_1t_4+2t_3t_5\right)\right]+3\left(t_1^2t_3-2t_1(t_2t_3+t_4^2)+4t_3t_5^2\right)\vphantom{\frac{a}{b}}\right\}
\end{aligned}\\
\begin{aligned}
\widetilde B_{44}=\frac{1}{36}&\left\{\vphantom{\frac{a}{b}}12k^4\left[((2r_3+r_5)(r_1+r_4+r_5))(t_1+t_3)\right]+2k^2\left[2(r_1+r_4+r_5)\left((t_1+4t_2)(t_1+t_3)\right.\right.\right.\\
&\left.\left.\left.+4(t_4+t_5)^2\right)+9(4r_3+r_5)t_1t_3-24r_7\left(t_1t_4-t_3t_5\right)\right] \vphantom{\frac{a}{b}}\right.\\
&\left.+3\left(t_1^2t_3+4t_1(t_2t_3+t_4^2+4t_3t_5^2)\right)\right\}
\end{aligned}
\end{array}
\end{equation*}
The determinant in eq.~\eqref{invs1dm} can be written as
\begin{align*}
\det\left(\widetilde b_{ij}^{\phi\chi}(1)\right)&=\frac{1}{9}\left((2r_3+r_5)(r_1+r_4+r_5)+r_7^2\right)\left((t_1+t_2)(t_1+t_3)+(t_4-2t_5)^2\right)\times\\
&\hspace{1cm}\times(k^2-m_+(1)^2)(k^2-m_-(1)^2) \numberthis \ ,
\end{align*}
where $m_\pm(1)^2$ are given by the following
\begin{align*}
m_{\pm}(1)^2=&-\frac{3}{4\left((2r_3+r_5)(r_1+r_4+r_5)+r_7^2\right)\left((t_1+t_2)(t_1+t_3)+(t_4-2t_5)^2\right)}\Bigg\{(r_1+r_4+r_5)\times\\
&\vphantom{\frac{a}{b}}\hspace{-.7cm}\times\left(t_1^2t_2+t_1(t_2t_3+t_4^2))+4t_2t_5^2\right)+(2r_3+r_5)\left(t_1^2t_3+t_1(t_2t_3+t_4^2))\right. \\
&\vphantom{\frac{A}{B}}\left.\hspace{-1cm}+4t_3t_5^2\right)-2r_7\left(t_1^2t_4-2(t_2t_3+(t_4-2t_5)t_4)t_5\right)\pm \Bigg[-4 \left((2r_3+r_5)\times\right.\\
&\left.\vphantom{\frac{a}{b}}\hspace{-.7cm}\times(r_1+r_4+r_5)+r_7^2\right)\left((t_1+t_2)(t_1+t_3)+(t_4-2t_5)^2\right)(t_2t_3+t_4^2)(t_1^2+4t_5^2) \\
&\vphantom{\frac{a}{b}}\hspace{-1cm}+\left[(r_1+r_4+r_5)\left(t_1^2t_2+t_1(t_2t_3+t_4^2))+4t_2t_5^2\right)+(2r_3+r_5)\times \right.\\
&\left.\vphantom{\frac{A}{B}}\hspace{-.7cm}\times\left(t_1^2t_3+t_1(t_2t_3+t_4^2))+4t_3t_5^2\right)-2r_7\left(t_1^2t_4\right.\right.\\
&\left.\left.\hspace{-1cm}-2(t_2t_3+(t_4-2t_5)t_4)t_5\right) \right]^2\vphantom{\frac{a}{b}}\Bigg]^\frac{1}{2} \Bigg\} \numberthis \ .
\end{align*}

\newpage

\subsection*{A.3: Spin-2}

\renewcommand{\theequation}{A.3.\arabic{equation}}
\setcounter{equation}{0}

The 9 operators corresponding to the tensor part of the theory are
\begin{align*}
\label{s2op}
&P^{\omega\omega}_{11}(2)_{CABKIJ}=\frac{4}{3}\, \Theta_{K(C}\Theta_{A)I}\Theta_{BJ}-\Theta_{CB}\Theta_{AI}\Theta_{JK}\ , \\
&P^{\omega\omega}_{12}(2)_{CABKIJ}=\frac{2}{3}\left(\epsilon_{ABD(J}\Theta_{K)C}-\epsilon_{BCD(J}\Theta_{K)A}\right)\Omega^D_I \ , \\
&P^{\omega\omega}_{21}(2)_{CABKIJ}=\frac{2}{3}\left(\epsilon_{IJL(B}\Theta_{C)K}-\epsilon_{JKL(B}\Theta_{C)I}\right)\Omega^L_A \ ,\\
&P^{\omega\omega}_{22}(2)_{CABKIJ}=2\,\Theta_{K(C}\Theta_{A)I}\Omega_{BJ}-\frac{2}{3}\, \Theta_{CB}\Omega_{AI}\Theta_{JK} \ , \   \\ 
&P^{\omega s}_{13}(2)_{CABIJ}=\frac{2\sqrt{2}}{3}\epsilon_{ADJ(B}\Theta_{C)I}\tilde k^D\ , \numberthis  \\
&P^{s\omega }_{31}(2)_{ABKIJ}=-\frac{2\sqrt{2}}{3}\epsilon_{ILB(J}\Theta_{K)A}\tilde k^L\ ,\\
&P^{\omega s}_{23}(2)_{CABIJ}=\sqrt{2}~\tilde k_B \left(\Theta_{CI}\Theta_{AJ}-\frac{1}{3}\Theta_{CA}\Theta_{IJ}\right) \ , \\
&P^{s\omega }_{32}(2)_{ABKIJ}=\sqrt{2}~\tilde k_J \left(\Theta_{KA}\Theta_{IB}-\frac{1}{3}\Theta_{KI}\Theta_{AB}\right) \ , \\
&P^{ss}_{33}(2)_{ABIJ}=\Theta_{AI}\Theta_{BJ}-\frac{1}{3}\Theta_{AB}\Theta_{IJ} \ .
\end{align*}

\noindent The coefficient matrix for the spin-2 sector is found to be
\begin{gather*}
\label{s2dm}
c_{ij}^{\phi\chi}(2)=\begin{blockarray}{cccc}
\scp\omega^-&\scp\omega^+&\scp s^+\\
\begin{block}{(ccc) c}
c_{11}&c_{12}&c_{13}&\scp\omega^- \\
c_{21}&c_{22}&c_{23}&\scp\omega^+ \\
c_{31}&c_{32}&c_{33}&\scp s^+ \\
\end{block} \numberthis
\end{blockarray}\ , \\
\renewcommand\arraystretch{1.2}
\arraycolsep=.4cm
\begin{array}{lll}
c_{11}= k^2 r_1+\frac{1}{2}t_1& c_{12}=k^2 r_8+t_5 & c_{13}=i \sqrt{2k^2}t_5  \\
c_{21}=-k^2 r_8-t_5 &c_{22}=k^2(2r_1-2r_3+r_4)+\frac{1}{2}t_1  & c_{23}=i \sqrt{\frac{k^2}{2}}t_1 \\
c_{31}=i \sqrt{2k^2}t_5  & c_{32}=-i \sqrt{\frac{k^2}{2}}t_1 & c_{33}=k^2(t_1-\lambda) 
\end{array}
\end{gather*}
Since the above is not a singular matrix, we can immediately calculate its inverse 
\begin{gather*}
\label{invs2dm}
\left(c_{ij}^{\phi\chi}(2)\right)^{-1}=\frac{k^2}{\det\left(c_{ij}^{\phi\chi}(2)\right)} \left(
\begin{array}{ccc}
C_{11}&C_{12}&C_{13}\\
C_{21}&C_{22}&C_{23}\\
C_{31}&C_{32}&C_{33}
\end{array} \right)\ , \numberthis \\
\vphantom{\frac{a}{b}}\\
\begin{aligned} 
&\begin{array}{ll} 
C_{11}=k^2(2r_1-2r_3+r_4)(t_1+\lambda)-\frac{1}{2}t_1\lambda &C_{12}=-k^2 r_8(t_1+\lambda)-\lambda t_5 \\
C_{21}=k^2 r_8(t_1+\lambda)+\lambda t_5 &C_{22}=\frac{1}{2}\left((k^2 r_1+t_1 )(t_1+\lambda)+2t_5^2\right) \\
C_{31}=i\sqrt{\frac{k^2}{2}}\left(r_8t_1-2(2r_1-2r_3+r_4)t_5\right) & C_{32}=\frac{i}{2}\sqrt{\frac{1}{2k^2}}\left(2k^2(r_1 t_1+2r_8t_5)+t_1^2+4t_5^2\right)\\
& \\
\end{array}\\
&\begin{array}{l}
C_{13}=i\sqrt{\frac{k^2}{2}}\left(r_8t_1-2(2r_1-2r_3+r_4)t_5\right) \\
C_{23}=-\frac{i}{2}\sqrt{\frac{1}{2k^2}}\left(2k^2(r_1 t_1+2r_8t_5)+t_1^2+4t_5^2\right) \\
C_{33}=k^2\left(r_1(2r_1-2r_3+r_4)+r_8^2\right)+\frac{1}{2}(3r_1-2r_3+r_4)t_1+2r_8t_5+\frac{1}{k^2}\left(\frac{1}{4}t_1^2+t_5^2\right)
\end{array}
\end{aligned}
\end{gather*}

The determinant of the matrix $c_{ij}^{\phi\chi}(2)$ reads
\begin{equation}
\det\left(c_{ij}^{\phi\chi}(2)\right)=\left(r_1(2r_1-2r_3+r_4)+r_8^2\right)(t_1+\lambda)k^2(k^2-m_+(2)^2)(k^2-m_-(2)^2) \ ,
\end{equation}
with $m_{\pm}(2)^2$ given by
\begin{align*}
m_{\pm}(2)^2&=\frac{1}{4\left(r_1(2r_1-2r_3+r_4)+r_8^2\right)(t_1+\lambda)}\Bigg\{-(2r_1-2r_3+r_4)\left(t_1^2+4 t_5^2\right)\\
&\hspace{-1cm}-(3r_1-2r_3+r_4)t_1\lambda+4r_8 t_5\lambda
\pm \Bigg[-4\left(r_1(2r_1-2r_3+r_4)+r_8^2\right)(t_1^2+t_5^2)(t_1+\lambda)\lambda \\
&\hspace{-1cm}+\left[(2r_1-2r_3+r_4)\left(t_1^2+t_5^2\right)+\left( (3r_1-2r_3+r_4)t_1+4r_8 t_5\right)\lambda\right]^2 \Bigg]^{\frac{1}{2}} \Bigg\} \numberthis \ ,
\end{align*}
where once again, we require the masses to be positive. 

Like in the scalar sector of the theory, it is very convenient to write the inverse coefficient matrix~\eqref{invs2dm} as
\begin{align}
\left(c_{ij}^{\phi\chi}(2)\right)^{-1}=&-\frac{1}{\lambda k^2}\left(\begin{array}{ccc}
0&0&0\\
0&-2k^2&i \sqrt{2k^2}\\
0&-i \sqrt{2k^2}&-1
\end{array}\right)  \nonumber \\
\vphantom{\frac{a}{b}}\nonumber \\
&\hspace{-2cm}+\frac{2}{t_1^2+4t_5^2}
\left(\begin{array}{ccc}
t_1&-2t_5&0\\
2t_5&t_1&0\\
0&0&\lambda^{-1}\left((3r_1-2r_3+r_4)t_1+8r_8 t_5\right)
\end{array}\right)\nonumber \\
\vphantom{\frac{a}{b}}\nonumber \\
&\hspace{-2cm}+\frac{1}{4\big((t_1+\lambda)(r_1(2r_1-2r_3+r_4)+ r_8^2)\big)\big(k^2-m_{2+}^{\ \ 2}\big)\big(k^2-m_{2-}^{\ \ 2}\big)}
\left(\begin{array}{ccc}
C_{11}&C_{12}&C_{13}\\
C_{21}&C_{22}&C_{23}\\
C_{31}&C_{32}&C_{33}
\end{array} \right) \numberthis \ .
\end{align}
The matrix elements $C_{ij}$ can be found above in eq.~\eqref{invs2dm}.

\newpage

\renewcommand{\theequation}{B.\arabic{equation}}
\setcounter{equation}{0}

\section*{Appendix B}
\label{app:derivproj}

\renewcommand{\theequation}{B.\arabic{equation}}
\setcounter{equation}{0}

In an attempt to make this article as self-contained as possible, we would like to give some details on the way the projectors used to decompose the theory into spin sectors are obtained. The operators are classified into two categories. The first contains the ``diagonal'' projectors $P^{\phi\phi}_{ii}(J)$, which correspond to the decomposition of the fields into irreducible representations of the three-dimensional rotations group. Their derivation amounts to addition of angular momenta, since with respect to $SO(3)$
\begin{equation*} 
\omega_{CAB}\rightarrow \  2^-\oplus 2^+\oplus 1^-\oplus 1^-\oplus 1^+\oplus 1^+\oplus 0^-\oplus 0^+ \ , \ \ \ h_{AB}\rightarrow \ 2^+\oplus 1^-\oplus 1^-\oplus 1^+\oplus 0^+\oplus 0^+ \ .  
\end{equation*}
In terms of $\Theta$ and $\Omega$, this decomposition of the fields can be written in covariant form as
\begin{align*}
\omega_{CAB}&=\left[\underbrace{\frac{4}{3}\, \Theta_{K(C}\Theta_{A)I}\Theta_{BJ}-\Theta_{CB}\Theta_{AI}\Theta_{JK}}_{P^{\omega\omega}_{11}(2)}+\underbrace{2\,\Theta_{K(C}\Theta_{A)I}\Omega_{BJ}-\frac{2}{3}\, \Theta_{CB}\Omega_{AI}\Theta_{JK}}_{P^{\omega\omega}_{22}(2)}+\underbrace{\vphantom{\frac{a}{b}}\Theta_{CB}\Theta_{AI}\Theta_{JK}}_{P^{\omega\omega}_{11}(1)}\right.\\
&\left.\hspace{1cm}+\underbrace{\vphantom{\frac{a}{b}}2~\Omega_{CB}\Theta_{AI}\Omega_{JK}}_{P^{\omega\omega}_{22}(1)}+\underbrace{\vphantom{\frac{a}{b}}\Theta_{CK}\Theta_{AI}\Omega_{BJ}+\Theta_{AK}\Omega_{BI}\Theta_{CJ}}_{P^{\omega\omega}_{55}(1)}+\underbrace{\vphantom{\frac{a}{b}}\Omega_{CK}\Theta_{AI}\Theta_{BJ}}_{P^{\omega\omega}_{66}(1)}\right. \\
&\left.\hspace{1cm}+\underbrace{\frac{1}{3} \Theta_{CK}\Theta_{AI}\Theta_{BJ}+\frac{2}{3} \Theta_{AK}\Theta_{BI}\Theta_{CJ}}_{P^{\omega\omega}_{11}(0)}+\underbrace{\frac{2}{3}\Theta_{BC}\Omega_{AI}\Theta_{JK}}_{P^{\omega\omega}_{22}(0)}\right]\omega^{KIJ}\numberthis \ ,
\end{align*}
and 
\begin{align*}
h_{AB}=&\left[\underbrace{\Theta_{AI}\Theta_{BJ}-\frac{1}{3}\Theta_{AB}\Theta_{IJ}}_{P^{ss}_{33}(2)}+\underbrace{\vphantom{\frac{a}{b}}2~\Theta_{AI}\Omega_{BJ} }_{P^{ss}_{33}(1)}+\underbrace{\vphantom{\frac{a}{b}}2~\Theta_{AI}\Omega_{BJ}}_{P^{aa}_{44}(1)}\right.\\
&\hspace{6cm}\left.+\underbrace{\vphantom{\frac{a}{b}}\Theta_{AI}\Theta_{BJ}}_{P^{aa}_{77}(1)}+\underbrace{\frac{1}{3}\Theta_{AB}\Theta_{IJ}
}_{P^{ss}_{33}(0)}+\underbrace{\vphantom{\frac{a}{b}}\Omega_{AB}\Omega_{IJ} }_{P^{ss}_{44}(0)}\right]h^{IJ}\numberthis \ .
\end{align*}

The second category contains the ``off-diagonal'' operators $P^{\phi\chi}_{ij}(J), \ \text{with} \ i\neq j$; they implement mappings between the same spin subspaces of the fields. They connect states with the same spin and same parity, as well as states with the same spin but different parity if the totally antisymmetric tensor is present. 

Consider the following mixing term between the symmetric part of the tetrad and the connection that contributes only to the scalar part of the theory
 \begin{equation}
k^B\eta^{CA}\eta^{DE}\omega_{CAB}\, s_{DE} \ .
\end{equation}
We wish to find the off-diagonal projectors that link the $J^P=0^+$ component of connection (projected out by $P^{\omega\omega}_{22}(0)$) to one of the $J^P=0^+$ components of the tetrad, for example $P^{ss}_{33}(0)$. Plugging the expressions for the operators from eq. \eqref{s0op} into the above, we find after some algebra that the mixing operators are proportional to
\begin{equation}
P^{\omega s}_{23}(0)_{CABIJ}=c(k)~k^B \Theta^{CA}\Theta^{IJ} \ \ \ \text{and} \ \ \ P^{s \omega }_{32}(0)_{ABKIJ}=c(k)~k^J \Theta^{KI}\Theta^{AB} \ .
\end{equation}
Here $c(k)$ is a coefficient that depends on momentum and is determined from the orthogonality relations~\eqref{orths}. In particular, for these operators we have
\begin{align} 
&P^{\omega s}_{23}(0)_{CABDE}\, P^{s \omega}_{32}(0)^{DE}_{\ \ \ KIJ}=P^{\omega\omega}_{22}(0)_{CABKIJ} \ ,\\
&P^{s \omega}_{32}(0)_{ABDEF}\,P^{\omega s}_{23}(0)^{DEF}_{\ \ \ \ IJ}=P^{ss}_{33}(0)_{ABIJ} \ ,
\end{align}
so we immediately find
\begin{equation}
c(k)=\frac{1}{3}\sqrt{\frac{2}{k^2}} \ .
\end{equation}

The construction of operators that are capable of handling terms that contain the totally antisymmetric symbol follows pretty much the same reasoning as in the previous example. A term like $\epsilon^{ABCD}a_{AB}a_{CD}$, mixes the $J^P=1^- \ (P^{aa}_{44}(1))$ with the $J^P=1^+\ (P^{aa}_{77}(1))$ states of the tetrad excitation. A straightforward computation reveals that the corresponding projectors read
\begin{equation}
P^{aa}_{47}(1)_{ABIJ}= c\, \epsilon_{AIJL}\Omega_B^L \ \ \ \text{and} \ \ \ \ P^{aa}_{74}(1)_{ABIJ}=c'\, \epsilon_{IABD}\Omega_J^D \ ,
\end{equation}
where in this case it is necessary to introduce two normalisation coefficients $c \ \text{and} \ c
'$, that do not depend on momentum. The orthogonality relations read
\begin{align}
&P^{aa}_{47}(1)_{ABCD}\,P^{aa}_{74}(1)^{CD}_{\ \ \ IJ}= P^{aa}_{44}(1)_{ABIJ} \ , \\
& P^{aa}_{74}(1)_{ABCD}\,P^{aa}_{47}(1)^{CD}_{\ \ \ IJ}= P^{aa}_{77}(1)_{ABIJ} \ , 
\end{align}
and in order for them to hold, we are required to set $c=-c'=1$. The fact that the projectors involving the totally antisymmetric tensor differ in sign is something that holds for all operators that connect states with opposite parities.   

Let us close this Appendix with a technical remark. Terms that contain the totally antisymmetric tensor are responsible for the appearance of mixing between states with (same spin but) different parity. Obviously, they must not affect the mixing of states with same parity. It is indeed easy to show explicitly that their contribution vanishes by using the Schouten identity 
\begin{equation} 
\epsilon^{ABCD}k^E+\epsilon^{BCDE}k^A+\epsilon^{CDEA}k^B+\epsilon^{DEAB}k^C+\epsilon^{EABC}k^D=0 \ .
\end{equation}

\section*{Appendix C}
\label{app:altern}
\renewcommand{\theequation}{C.\arabic{equation}}
\setcounter{equation}{0}

In this Appendix we present an alternative way to determine the conditions for absence of ghosts and tachyons. In this approach, we do not need to invert the wave operator at any time, but we have to sacrifice covariance and work in terms of the components of the fields. This makes the calculations very lengthy and difficult to keep track of. These reasons make the method presented here much less appealing than the one we followed in the main text. However, it can be used as a cross-check for our results, since the restrictions on the coefficients are the same as the ones we obtained by analysing the behaviour of the propagators, see eqs.~\eqref{conditions-full-s0}-\eqref{conditions-full-s2}.

In what follows, we will discuss in detail only the spin-1 sector of the theory, the reason being that there is no mixing between tetrad and connection. This makes the calculations and the expressions considerably simpler with respect spin-0 and spin-2 sectors. The generalisation of this method in the presence of (kinetic) mixing between the different fields is straightforward, something we will illustrate at the end of this Appendix. 

As a first step, we apply the weak field approximation to the theory and write its linearised form around Minkowski background in momentum space. We can now use the projection operators (given in Appendix A) to decompose the resulting action for the perturbations into independent spin sub-spaces. Since we have already derived the coefficient matrices, we will use them here as well. This is purely for convenience, since one does not need to use the full set of projectors to proceed. Once the theory is broken into separate spin sectors, upon using the diagonal operators, the coefficient matrices can be derived by inspection of the action.

The invariance of the initial theory under general coordinate and local Lorentz transformations has been passed down on its linearised version as well. As a result, certain degrees of freedom are pure gauge. Since they have no physical significance, they can be removed by imposing appropriate gauge conditions. For example, we found it helpful to follow Neville~\cite{Neville:1978bk} and choose
\begin{equation}
\label{gauge-fix}
\partial^As_{AB}=0 \ \ \ \text{and} \ \ \ a_{AB}=0  \ .
\end{equation}
The gauge-fixed part of the action that describes the dynamics of spin-1 particles reads 
\begin{equation}
\label{spin-1-alt}
S_{2}(\text{spin-1})=\int d^4x \sum_{i,j} \widetilde b^{\omega\omega}_{ij}(1)\,\omega_{CAB}^* \,P^{\omega\omega}_{ij}(1)^{CABKIJ}\,\omega_{KIJ} +\ \text{source terms} \ , 
\end{equation}
where summation over Lorentz indices is understood. Here $\omega_{CAB}\equiv\omega_{CAB}(k)$ is the connection field in momentum space and $\omega(k)_{CAB}^*=\omega(-k)_{CAB}$. The explicit expressions for the projection operators $P^{\omega\omega}_{ij}(1)$ and the corresponding coefficient matrix $\widetilde b^{\omega\omega}_{ij}(1)$ were given in eqs.~\eqref{s1op} and~\eqref{s1dm} respectively. 

At this point we move to the rest frame of the particles, since there the four-momentum is $k^\mu=(k^0,0,0,0)$ and the expressions simplify a lot. In terms of the components of $\omega$, the quadratic form in the above action can be written compactly as
\begin{equation}
\label{mat-not-1}
\sum_{i,j} \widetilde b^{\omega\omega}_{ij}(1)\,\omega_{CAB}^* \,P^{\omega\omega}_{ij}(1)^{CABKIJ}\,\omega_{KIJ} =\sum_{i,j}\phi'_i\, b'_{ij}(1)\, \phi'_j \ , 
\end{equation}
where $\phi'_i$ contains the components that are involved in the vector part of the theory and it reads
\begin{equation}
\phi'_i=\left(\begin{array}{c}
\left(\omega^{113}+\omega^{223}\right)\\
\left(\omega^{212}+\omega^{313}\right)\\
\left(\omega^{112}-\omega^{323}\right)\\
\omega^{010}\\
\omega^{020}\\
\omega^{030}\\
\left(\omega^{120}-\omega^{210}\right)\\
\left(\omega^{130}-\omega^{310}\right)\\
\left(\omega^{230}-\omega^{320}\right)\\
\omega^{012}\\
\omega^{013}\\
\omega^{023}
\end{array}
\right) \ .
\end{equation}
The (non-degenerate) $12\times 12$ matrix $b'_{ij}(1)$ is given by
\begin{equation*}
\label{bprime-comps}
b'_{ij}(1)=
\begin{tikzpicture}[baseline=-0.5ex]
\matrix [matrix of math nodes,left delimiter=(,right delimiter=)] (b')
   { b'_{11}&b'_{12}&b'_{13}&b'_{14}&b'_{15}&b'_{16}&b'_{17}&b'_{18}&b'_{19}&b'_{1,10}&b'_{1,11}&b'_{1,12}\\
b'_{21}&b'_{22}&b'_{23}&b'_{24}&b'_{25}&b'_{26}&b'_{27}&b'_{28}&b'_{29}&b'_{2,10}&b'_{2,11}&b'_{2,12}\\
b'_{31}&b'_{32}&b'_{33}&b'_{34}&b'_{35}&b'_{36}&b'_{37}&b'_{38}&b'_{39}&b'_{3,10}&b'_{3,11}&b'_{3,12}\\
b'_{41}&b'_{42}&b'_{43}&b'_{44}&b'_{45}&b'_{46}&b'_{47}&b'_{48}&b'_{49}&b'_{4,10}&b'_{4,11}&b'_{4,12}\\
b'_{51}&b'_{52}&b'_{53}&b'_{54}&b'_{55}&b'_{56}&b'_{57}&b'_{58}&b'_{59}&b'_{5,10}&b'_{5,11}&b'_{5,12}\\
b'_{61}&b'_{62}&b'_{63}&b'_{64}&b'_{65}&b'_{66}&b'_{67}&b'_{68}&b'_{69}&b'_{6,10}&b'_{6,11}&b'_{6,12}\\
b'_{71}&b'_{72}&b'_{73}&b'_{74}&b'_{75}&b'_{76}&b'_{77}&b'_{78}&b'_{79}&b'_{7,10}&b'_{7,11}&b'_{7,12}\\
b'_{81}&b'_{82}&b'_{83}&b'_{84}&b'_{85}&b'_{86}&b'_{87}&b'_{88}&b'_{89}&b'_{8,10}&b'_{8,11}&b'_{8,12}\\
b'_{91}&b'_{92}&b'_{93}&b'_{94}&b'_{95}&b'_{96}&b'_{97}&b'_{98}&b'_{99}&b'_{9,10}&b'_{9,11}&b'_{9,12}\\
b'_{10,1}&b'_{10,2}&b'_{10,3}&b'_{10,4}&b'_{10,5}&b'_{10,6}&b'_{10,7}&b'_{10,8}&b'_{10,9}&b'_{10,10}&b'_{10,11}&b'_{10,12}\\
b'_{11,1}&b'_{11,2}&b'_{11,3}&b'_{11,4}&b'_{11,5}&b'_{11,6}&b'_{11,7}&b'_{11,8}&b'_{11,9}&b'_{11,10}&b'_{11,11}&b'_{11,12}\\
b'_{12,1}&b'_{12,2}&b'_{12,3}&b'_{12,4}&b'_{12,5}&b'_{12,6}&b'_{12,7}&b'_{12,8}&b'_{12,9}&b'_{12,10}&b'_{12,11}&b'_{12,12}\\
        };  
\draw[color=black,style=thick,dashed] (b'-1-1.north west) -- (b'-1-6.north east) -- (b'-6-6.south east) -- (b'-6-1.south west) -- (b'-1-1.north west);
 \draw[color=black,style=thick,dashed] (b'-7-7.north west) -- (b'-7-12.north east) -- (b'-12-12.south east) -- (b'-12-7.south west) -- (b'-7-7.north west);
    \end{tikzpicture}
\numberthis \ ,
\end{equation*}
\begin{align*}
\renewcommand\arraystretch{1.2}
\arraycolsep=.4cm
&\begin{array}{lll}
b'_{11}=-\left[(k^0)^2(r_1+r_4+r_5)+\frac{1}{6}(t_1+4t_3)\right]&&b'_{12}=0\\
b'_{21}=0&&b'_{22}=-\left[(k^0)^2(r_1+r_4+r_5)+\frac{1}{6}(t_1+4t_3)\right]\\
b'_{31}=0&&b'_{32}=0\\
b'_{41}=0&&b'_{42}=-\frac{1}{3}(t_1-2t_3)\\
b'_{51}=0&&b'_{52}=0\\
b'_{61}=\frac{1}{3}(t_1-2t_3)&&b'_{62}=0\\
b'_{71}=(k^0)^2r_7-\frac{1}{3}(2t_4-t_5)&&b'_{72}=0\\
b'_{81}=0&&b'_{82}=0\\
b'_{91}=0&&b'_{92}=-\left[(k^0)^2r_7-\frac{1}{3}(2t_4-t_5)\right]\\
b'_{10,1}=-\frac{2}{3}(t_4+t_5)&&b'_{10,2}=0\\
b'_{11,1}=0&&b'_{11,2}=0\\
b'_{12,1}=0&&b'_{12,2}=\frac{2}{3}(t_4+t_5)\\
& \\
\end{array}\\
&\begin{array}{lll}
b'_{13}=0&&b'_{14}=0\\
b'_{23}=0&&b'_{24}=-\frac{1}{3}(t_1-2t_3)\\
b'_{33}=-\left[(k^0)^2(r_1+r_4+r_5)+\frac{1}{6}(t_1+4t_3)\right]&&b'_{34}=0\\
b'_{43}=0&&b'_{44}=-\frac{2}{3}(t_1+t_3)\\
b'_{53}=\frac{1}{3}(t_1-2t_3)&&b'_{54}=0\\
b'_{63}=0&&b'_{64}=0\\
b'_{73}=0&&b'_{74}=0\\
b'_{83}=-\left[(k^0)^2r_7-\frac{1}{3}(2t_4-t_5)\right]&&b'_{84}=0\\
b'_{93}=0&&b'_{94}=-\frac{2}{3}(t_4+t_5)\\
b'_{10,3}=0&&b'_{10,4}=0\\
b'_{11,3}=\frac{2}{3}(t_4+t_5)&&b'_{11,4}=0\\
b'_{12,3}=0&&b'_{12,4}=-\frac{2}{3}(t_4-2t_5)\\
& \\
\end{array}\\
&\begin{array}{lllll}
b'_{15}=0&&b'_{16}=\frac{1}{3}(t_1-2t_3)&&b'_{17}=(k^0)^2r_7-\frac{1}{3}(2t_4-t_5)\\
b'_{25}=0&&b'_{26}=0&&b'_{27}=0\\
b'_{35}=\frac{1}{3}(t_1-2t_3)&&b'_{36}=0&&b'_{37}=0\\
b'_{45}=0&&b'_{46}=0&&b'_{47}=0\\
b'_{55}=-\frac{2}{3}(t_1+t_3)&&b'_{56}=0&&b'_{57}=0\\
b'_{65}=0&&b'_{66}=-\frac{2}{3}(t_1+t_3)&&b'_{67}=\frac{2}{3}(t_4+t_5)\\
b'_{75}=0&&b'_{76}=\frac{2}{3}(t_4+t_5)&&b'_{77}=(k^0)^2(2r_3+r_5)+\frac{1}{6}(t_1+4t_2)\\
b'_{85}=\frac{2}{3}(t_4+t_5)&&b'_{86}=0&&b'_{87}=0\\
b'_{95}=0&&b'_{96}=0&&b'_{97}=0\\
b'_{10,5}=0&&b'_{10,6}=-\frac{2}{3}(t_4-2t_5)&&b'_{10,7}=-\frac{1}{3}(t_1-2t_2)\\
b'_{11,5}=\frac{2}{3}(t_4-2t_5)&&b'_{11,6}=0&&b'_{11,7}=0\\
b'_{12,5}=0&&b'_{12,6}=0&&b'_{12,7}=0\\
& \\
b'_{18}=0&&&&b'_{19}=0\\
b'_{28}=0&&&&b'_{29}=-\left[(k^0)^2r_7-\frac{1}{3}(2t_4-t_5)\right]\\
b'_{38}=-\left[(k^0)^2r_7-\frac{1}{3}(2t_4-t_5)\right]&&&&b'_{39}=0\\
b'_{48}=0&&&&b'_{49}=-\frac{2}{3}(t_4+t_5)\\
b'_{58}=\frac{2}{3}(t_4+t_5-)&&&&b'_{59}=0\\
b'_{68}=0&&&&b'_{69}=0\\
b'_{78}=0&&&&b'_{79}=0\\
b'_{88}=(k^0)^2(2r_3+r_5)+\frac{1}{6}(t_1+4t_2)&&&&b'_{89}=0\\
b'_{98}=0&&&&b'_{99}=(k^0)^2(2r_3+r_5)+\frac{1}{6}(t_1+4t_2)\\
b'_{10,8}=0&&&&b'_{10,9}=0\\
b'_{11,8}=-\frac{1}{3}(t_1-2t_2)&&&&b'_{11,9}=0\\
b'_{12,8}=0&&&&b'_{12,9}=-\frac{1}{3}(t_1-2t_2)\\
& \\
\end{array}\\
&\begin{array}{lllllll}
b'_{1,10}=-\frac{2}{3}(t_4+t_5)&&b'_{1,11}=0&&b'_{1,12}=0\\
b'_{2,10}=0&&b'_{2,11}=0&&b'_{2,12}=\frac{2}{3}(t_4+t_5)\\
b'_{3,10}=0&&b'_{3,11}=\frac{2}{3}(t_4+t_5)&&b'_{3,12}=0\\
b'_{4,10}=0&&b'_{4,11}=0&&b'_{4,12}=-\frac{2}{3}(t_4-2t_5)\\
b'_{5,10}=0&&b'_{5,11}=\frac{2}{3}(t_4-2t_5)&&b'_{5,12}=0\\
b'_{6,10}=-\frac{2}{3}(t_4-2t_5)&&b'_{6,11}=0&&b'_{6,12}=0\\
b'_{7,10}=-\frac{1}{3}(t_1-2t_2)&&b'_{7,11}=0&&b'_{7,12}=0\\
b'_{8,10}=0&&b'_{8,11}=-\frac{1}{3}(t_1-2t_2)&&b'_{8,12}=0\\
b'_{9,10}=0&&b'_{9,11}=0&&b'_{9,12}=-\frac{1}{3}(t_1-2t_2)\\
b'_{10,10}=\frac{2}{3}(t_1+t_2)&&b'_{10,11}=0&&b'_{10,12}=0\\
b'_{11,10}=0&&b'_{11,11}=\frac{2}{3}(t_1+t_2)&&b'_{11,12}=0\\
b'_{12,10}=0&&b'_{12,11}=0&&b'_{12,12}=\frac{2}{3}(t_1+t_2)\\
& \\
\end{array}
\end{align*}
Once again, we have arranged the matrix elements so that the $1^-(1^+)$ sector is described by the $6\times6$ top-left (bottom-right) sub-matrix that is outlined in~\eqref{bprime-comps}. If there are no parity-odd pieces present in the theory, then the matrix becomes block-diagonal. Vector and pseudovector sectors decouple completely, as they should.

Inspection of the matrix $b'_{ij}(1)$ reveals that out of the 18 components of the connection that are involved, the dynamical degrees of freedom associated to the spin-1 part of the action appear in the following six combinations only\footnote{This is of course due to the decomposition of the connection perturbations with the help of the projection operators.}
\begin{eqnarray}
&\text{vector}\ (1^-):& \ \ \ \omega^{113}+\omega^{223} \ ,\ \ \ \omega^{221}+\omega^{331}\ ,\ \ \ \omega^{112}+\omega^{332} \ , \\
&\text{pseudovector}\ (1^+):&\omega^{120}-\omega^{210} \ ,\ \ \ \omega^{130}-\omega^{310}\ ,\ \ \ \omega^{230}-\omega^{320} \ .
\end{eqnarray}
Indeed, this is exactly what we expected. 

Meanwhile, the quantities $\omega^{010},\omega^{020},\omega^{030},\omega^{012},\omega^{013},\omega^{023},$ appear with no time derivatives in the \emph{full action} (i.e. they are not multiplied by $k^0$ in the momentum representation that we use here) and their corresponding source terms in the rest frame vanish by virtue of $\sigma^{0AB}=0 \ .$ Since these components enter the action quadratically, we can integrate them out in the standard way (see for example~\cite{Rubakov:2008nh}): after varying the action with respect to these fields, we use the equations of motion to express them in terms of the dynamical components. We can then eliminate them by substituting the results back in the action. After a straightforward calculation we find that the spin-1 action given above in  eq.~\eqref{spin-1-alt} and~\eqref{mat-not-1}, can be written conveniently as sum of a kinetic $(K_{ij})$ and a mass $(M_{ij})$ matrix
\begin{equation}
\label{split}
\sum_{i,j}\phi_{i}\left(K_{ij}-M_{ij}\right)\phi_{i} \ , 
\end{equation}
where $\phi_{i}$ contains only the dynamical fields:
\begin{equation}
\phi_{i}=\left(\begin{array}{c}
\left(\omega^{113}+\omega^{223}\right)\\
\left(\omega^{212}+\omega^{313}\right)\\
\left(\omega^{112}-\omega^{323}\right)\\
\left(\omega^{120}-\omega^{210}\right)\\
\left(\omega^{130}-\omega^{310}\right)\\
\left(\omega^{230}-\omega^{320}\right)\\
\end{array}\right) \ .
\end{equation}
The kinetic matrix $K_{ij}$ is given by
\begin{equation}
\label{kin-mat-1}
K_{ij}=\left(k^0\right)^2\left(\begin{array}{cccccc}
K_{11}&K_{12}&K_{13}&K_{14}&K_{15}&K_{16}\\
K_{21}&K_{22}&K_{23}&K_{24}&K_{25}&K_{26}\\
K_{31}&K_{32}&K_{33}&K_{34}&K_{35}&K_{36}\\
K_{41}&K_{42}&K_{43}&K_{44}&K_{45}&K_{46}\\
K_{51}&K_{52}&K_{53}&K_{54}&K_{55}&K_{56}\\
K_{61}&K_{62}&K_{63}&K_{64}&K_{65}&K_{66}
\end{array}\right) \ ,
\end{equation}
\begin{align*}
&\begin{array}{lllll}
K_{11}=-(r_1+r_4+r_5)&&K_{12}=0&&K_{13}=0\\
K_{21}=0&&K_{22}=-(r_1+r_4+r_5)&&K_{23}=0\\
K_{31}=0&&K_{32}=0&&K_{33}=-(r_1+r_4+r_5)\\
K_{41}=r_7&&K_{42}=0&&K_{43}=0\\
K_{51}=0&&K_{52}=0&&K_{53}=-r_7\\
K_{61}=0&&K_{62}=-r_7&&K_{63}=0\\
\\
K_{14}=r_7&&K_{15}=0&&K_{16}=0\\
K_{24}=0&&K_{25}=0&&K_{26}=-r_7\\
K_{34}=0&&K_{35}=-r_7&&K_{36}=0\\
K_{44}=2r_3+r_5&&K_{45}=0&&K_{46}=0\\
K_{54}=0&&K_{55}=2r_3+r_5&&K_{56}=0\\
K_{64}=0&&K_{65}=0&&K_{66}=2r_3+r_5\\
\end{array}
\end{align*}

The mass matrix $M_{ij}$ reads
\begin{equation}
\label{mass-mat-1}
M_{ij}=-\frac{3}{2\left((t_1+t_2)(t_1+t_3)+(t_4-2t_5)^2\right)}\left(\begin{array}{cccccc}
M_{11}&M_{12}&M_{13}&M_{14}&M_{15}&M_{16}\\
M_{21}&M_{22}&M_{23}&M_{24}&M_{25}&M_{26}\\
M_{31}&M_{32}&M_{33}&M_{34}&M_{35}&M_{36}\\
M_{41}&M_{42}&M_{43}&M_{44}&M_{45}&M_{46}\\
M_{51}&M_{52}&M_{53}&M_{54}&M_{55}&M_{56}\\
M_{61}&M_{62}&M_{63}&M_{64}&M_{65}&M_{66}
\end{array}\right) \ ,
\end{equation}
{\small
\begin{align*}	
\hspace{-.9cm}
&\begin{array}{ll}
M_{11}=-t_3(t_1^2+4t_5^2)-t_1(t_2t_3+t_4^2)&M_{12}=0\\
M_{21}=0&M_{22}=-t_3(t_1^2+4t_5^2)-t_1(t_2t_3+t_4^2)\\
M_{31}=0&M_{32}=0\\
M_{41}=-t_1^2t_4+2t_5(t_2t_3+(t_4-2t_5)t_4)&M_{42}=0\\
M_{51}=0&M_{52}=0\\
M_{61}=0&M_{62}=t_1^2t_4-2t_5(t_2t_3+(t_4-2t_5)t_4)\\
\\
M_{13}=0&M_{14}=-t_1^2t_4+2t_5(t_2t_3+(t_4-2t_5)t_4)\\
M_{23}=0&M_{24}=0\\
M_{33}=-t_3(t_1^2+4t_5^2)-t_1(t_2t_3+t_4^2)&M_{34}=0\\
M_{43}=0&M_{44}=t_2(t_1^2+4t_5^2)+t_1(t_2t_3+t_4^2)\\
M_{53}=t_1^2t_4-2t_5(t_2t_3+(t_4-2t_5)t_4)&M_{54}=0\\
M_{63}=0&M_{64}=0\\
\\
M_{15}=0&M_{16}=0\\
M_{25}=0&M_{26}=t_1^2t_4-2t_5(t_2t_3+(t_4-2t_5)t_4)\\
M_{35}=t_1^2t_4-2t_5(t_2t_3+(t_4-2t_5)t_4)&M_{36}=0\\
M_{45}=0&M_{46}=0\\
M_{55}=t_2(t_1^2+4t_5^2)+t_1(t_2t_3+t_4^2)&M_{56}=0\\
M_{65}=0&M_{66}=t_2(t_1^2+4t_5^2)+t_1(t_2t_3+t_4^2)
\end{array}
\end{align*}}

In the way we have written the action, we can now immediately determine the conditions for absence of ghost and tachyons: the kinetic and mass matrices must be positive-definite respectively. A necessary and sufficient condition for that to hold is to require that the principal minors of these matrices are positive. We first apply this criterion on the kinetic matrix~\eqref{kin-mat-1} to find
\begin{equation}
(r_1+r_4+r_5)<0 \ \ \ \text{and} \ \ \ (r_1+r_4+r_5)(2r_3+r_5)<-r_7^2\ .
\end{equation}
Since $r_7^2>0$, the above implies
\begin{equation}
2r_3+r_5 >-\frac{r_7^2}{r_1+r_4+r_5}  \ ,
\end{equation}
as expected and in accordance with the conditions in eqs.~\eqref{conditions-sn-s0p}-\eqref{conditions-sn-s2m} for absence of ghosts in the parity-preserving theory (in which $r_7=0$). Next we move to the mass matrix~\eqref{mass-mat-1}, where we see that in order to have positive masses, the following constraints on the parameters must be imposed
\begin{equation}
\begin{aligned}
&(t_1+t_2)(t_1+t_3)+(t_4-2t_5)^2<0\ , \ \ \  t_2 t_3+t_4^2>0 \ , \\ &t_1^2+4t_5^2>0\ ,\ \ \ t_3(t_1^2+4t_5^2)>-t_1(t_2t_3+t_4^2) \ . 
\end{aligned}
\end{equation}
Once again, in the absence of parity-odd invariants in the action, the above boil down to
\begin{equation}
t_1t_2(t_1+t_2)<0 \ \ \ \text{and} \ \ \ t_1 t_3(t_1+t_3)>0 \ .
\end{equation}
Notice that the conditions obtained this way are the same as the ones we found from analysing the propagators of the different spin sectors. 

Before concluding this Appendix, we want to shortly outline how the above methodology can be applied to the scalar and tensor perturbations of the theory. These sectors are a bit more involved with respect to the spin-1 sector that we studied in detail here. Due to the kinetic mixing between tetrad and connection excitations (terms containing one derivative), the splitting of the action into kinetic and mass contributions becomes more complicated. Meanwhile, unlike the case we demonstrated, all components involved in the spin-0 and spin-2 sectors are propagating degrees of freedom, so they cannot be integrated out. However, what one can do is use the equations of motion to decouple the fields by getting rid of their kinetic mixing. This is achieved by solving the equations of motion, say for the tetrad, and plugging the result back in the action. Since the solution involves contributions from source terms, the corresponding fields \emph{are not integrated out}. This procedure makes it possible to write the action in the form presented above in eq.~\eqref{split}, i.e. as sum of a kinetic matrix (two derivatives) and a mass matrix (no derivatives). Once again, the absence for ghost and tachyons is equivalent to the matrices being positive-definite. This requirement leads to the following constraints for the spin-0 sector
\begin{equation}
\begin{aligned}
&r_2<0\ ,\ \ \ 2r_2(r_1-r_3+2r_4)<-r_6^2\ ,\ \ \  r_1-r_3+2r_4>-\frac{r_6^2}{r_2} \ ,\\
&t_2(t_3-\lambda)+t_4^2>0\ ,\ \ \ \left(t_2 t_3+t_4^2\right)\lambda(t_3-\lambda)>0 \ .
\end{aligned}
\end{equation}
For the spin-2 sector we find
\begin{equation}
\begin{aligned}
&r_1<0\ ,\ \ \ r_1(2r_1-2r_3+r_4)<-r_8^2\ , \ \ \ 2r_1-2r_3+r_4>-\frac{r_8^2}{r_1} \ , \\
&t_1\lambda(t_1+\lambda)<0 \ , \ \ \ t_1(t_1+\lambda)+4t_5^2 >0 \ .
\end{aligned}
\end{equation}

\end{appendix}

\newpage
\setcounter{page}{1}
\begin{center}

{\Large\bf Erratum:\\
The particle spectrum of  parity-violating Poincar\'e gravitational theory}

\vspace{.4cm}

{\large  Georgios K. Karananas}\\

\vspace{.4cm}

 {\it {Institut de Th\'eorie des Ph\'enom\`enes Physiques, \'Ecole Polytechnique F\'ed\'erale de Lausanne,\\
CH-1015 Lausanne, Switzerland}}\\

\vspace{.4cm}
\texttt{georgios.karananas@epfl.ch}\\

\vspace{.4cm}
February 2015
\end{center}

\vspace{.4cm}

\noindent In~\cite{Karananas:2014pxa} I analyzed the dynamics of the particle states of the Poincar\'e gravitational theory with parity-odd terms in the action. I arrived at the conclusion that the theory under investigation maybe free from ghosts in a particular region of the parameter space, determined by the
following set of inequalities that the parameters appearing in front of the kinetic terms satisfy
\begin{align}
\label{conditions-full1}
&r_1<0\ , \\ 
&r_2<0\ ,\\ 
&r_1-r_3+2r_4>-\frac{r_6^2}{2r_2} \ ,\\
\label{conditions-full2}
&r_1+r_4+r_5<0\ , \\ 
\label{conditions-full3} 
&2r_3+r_5>-\frac{r_7^2}{r_1+r_4+r_5}\ ,\\
\label{conditions-full4}
&2r_1-2r_3+r_4>-\frac{r_8^2}{r_1} \ .
\end{align}
As pointed out to me by James Nester after the paper was published in the journal, I overlooked the fact that when~\eqref{conditions-full2},~\eqref{conditions-full3} and~\eqref{conditions-full4} are summed, they yield
\begin{equation}
r_1>-\frac{r_7^2}{r_1+r_4+r_5}-\frac{r_8^2}{r_1}>0 \ ,
\end{equation}
which is in contradiction with~\eqref{conditions-full1}. As a result, I withdraw the claim that the theory is ghost-free once parity violating terms are taken into account. Notice however that there still exist boundaries of the extended parameter space where ghosts may not be present. 

\noindent On the other hand, the conditions I derived for the absence of tachyonic states remain unaffected.  

\section*{Acknowledgements}

\noindent I am very grateful to James Nester for bringing to my attention this contradiction.

\end{document}